%% file: main.tex
\documentclass[sigconf,nonacm]{acmart}

\AtBeginDocument{%
  }

\copyrightyear{2026}
\acmYear{2026}

\usepackage[dvipsnames]{xcolor}
\usepackage{xspace}
\usepackage{listings}
\usepackage{float}
\usepackage{hyperref}
\usepackage{enumitem}
\usepackage{subcaption}
\usepackage{dsfont}
\usepackage{tabularx}
\usepackage{minted}
\usepackage{multirow}
\usepackage[strings]{underscore}
\usepackage{makecell}
\usepackage{longtable}
\usepackage{pifont}
\usepackage{algorithm}
\usepackage{tcolorbox}
\usepackage[noend]{algpseudocode}
\usepackage{ifthen}

\usepackage{amsmath,amssymb}
\usepackage{amsfonts}
\usepackage{graphicx}
\usepackage{booktabs}
\usepackage[table]{xcolor}

\newif\ifcomment
\commenttrue
\newboolean{showcomments}
\setboolean{showcomments}{false}

\newcommand{\skr}[1]{{\ifthenelse{\boolean{showcomments}} {\color{blue}{#1}}{}}}
\newcommand{\yf}[1]{{\ifthenelse{\boolean{showcomments}} {\color{red}{#1}}{}}}
\newcommand{\tx}[1]{{\ifthenelse{\boolean{showcomments}} {\color{orange}{#1}}{}}}

\commentfalse
\setboolean{showcomments}{false}

\begin{document}

\title{SLAC: Access-Driven CPU-to-GPU Side-channel Attacks via System-Level Cache on Apple Silicon}

\author{Tianhong Xu}
\affiliation{%
  \institution{Northeastern University}
  \city{Boston}
  \country{USA}}
\email{xu.tianh@northeastern.edu}

\author{Saion K. Roy}
\affiliation{%
  \institution{Northeastern University}
  \city{Boston}
  \country{USA}}
\email{sai.roy@northeastern.edu}

\author{Ruyi Ding}
\affiliation{%
  \institution{Louisiana State University}
  \city{Baton Rouge}
  \country{USA}}
\email{ruyiding@lsu.edu}

\author{Aidong Adam Ding}
\affiliation{%
  \institution{Northeastern University}
  \city{Boston}
  \country{USA}}
\email{a.ding@northeastern.edu}

\author{Yunsi Fei}
\affiliation{%
  \institution{Northeastern University}
  \city{Boston}
  \country{USA}}
\email{y.fei@northeastern.edu}

\begin{abstract}
Modern heterogeneous System-on-Chip designs integrate CPU cores and a GPU that share a last-level cache (LLC) or system-level cache (SLC). This sharing exposes a new cross-domain attack surface, and existing attacks on integrated platforms either exploit coarse-grained cache-occupancy contention or require the adversary to co-reside on the GPU with the victim to obtain accurate timing measurements. In this work, we target Apple Silicon heterogeneous SoCs and discover that GPU memory accesses leave set-level footprints in the shared SLC, observable to an unprivileged CPU process. This keen observation enables the first fine-grained, access-driven, Prime+Probe-style CPU-to-GPU cache side-channel attacks against GPU workloads. We first reverse-engineer the Apple M1 SLC set-indexing functions and the interactions between local private caches and the SLC. Building on these findings, we construct the CPrime+CProbe SLC side-channel technique, which monitors GPU victim activity from the CPU at cache-set granularity. We then introduce an accelerated variant, GPrime+CProbe, in which an adversary leverages the GPU for faster SLC priming, yielding a 6.4× increase in the covert-channel throughput. Lastly, we demonstrate two end-to-end privacy attacks using the new side-channels: a graph-edge reconstruction attack on Graph Neural Networks (GNNs) that achieves 90\% edge accuracy across five datasets, and an LLM privacy attack that recovers input keywords with up to 94.8\% accuracy and model responses with up to 88.9\% accuracy across TinyLlama and GPT-2 Medium models. Our results reveal a new class of microarchitectural vulnerabilities in Apple Silicon and call for secure system cache designs for heterogeneous SoCs.
\end{abstract}



\maketitle

\input{txt/1.tex}

\input{txt/2.tex}

\input{txt/3.tex}
\input{txt/4.tex}

\input{txt/5.tex}
\input{txt/6.tex}

\begin{acks}
This work was supported in part by National Science Foundation under grant CNS-1916762 and IUCRC Center for Hardware and Embedded Systems Security and Trust (CHEST).
\end{acks}

\bibliographystyle{ACM-Reference-Format}
\bibliography{sample-base}

\input{txt/appendix}

\end{document}

%% file: txt/1.tex
\section{Introduction}

Modern GPUs are used extensively for privacy-sensitive workloads such as machine learning (ML) inference and cryptographic operations, making information leakage from these computations a pressing security concern. Heterogeneous computing systems integrate CPU and GPU on the same system-on-chip (SoC), and their shared microarchitectural resources create a new cross-domain path to confidentiality and privacy breaches of GPU workloads. Realizing such microarchitectural attacks, however, is nontrivial and highly platform-dependent: \textit{the success hinges on which microarchitectural resources are shared between CPU and GPU, how those resources are accessed or contended by the two sides, and whether GPU activities can be reliably observed from the CPU side.}

\begin{figure}[h]
    \centering
    \includegraphics[width=0.9\linewidth]{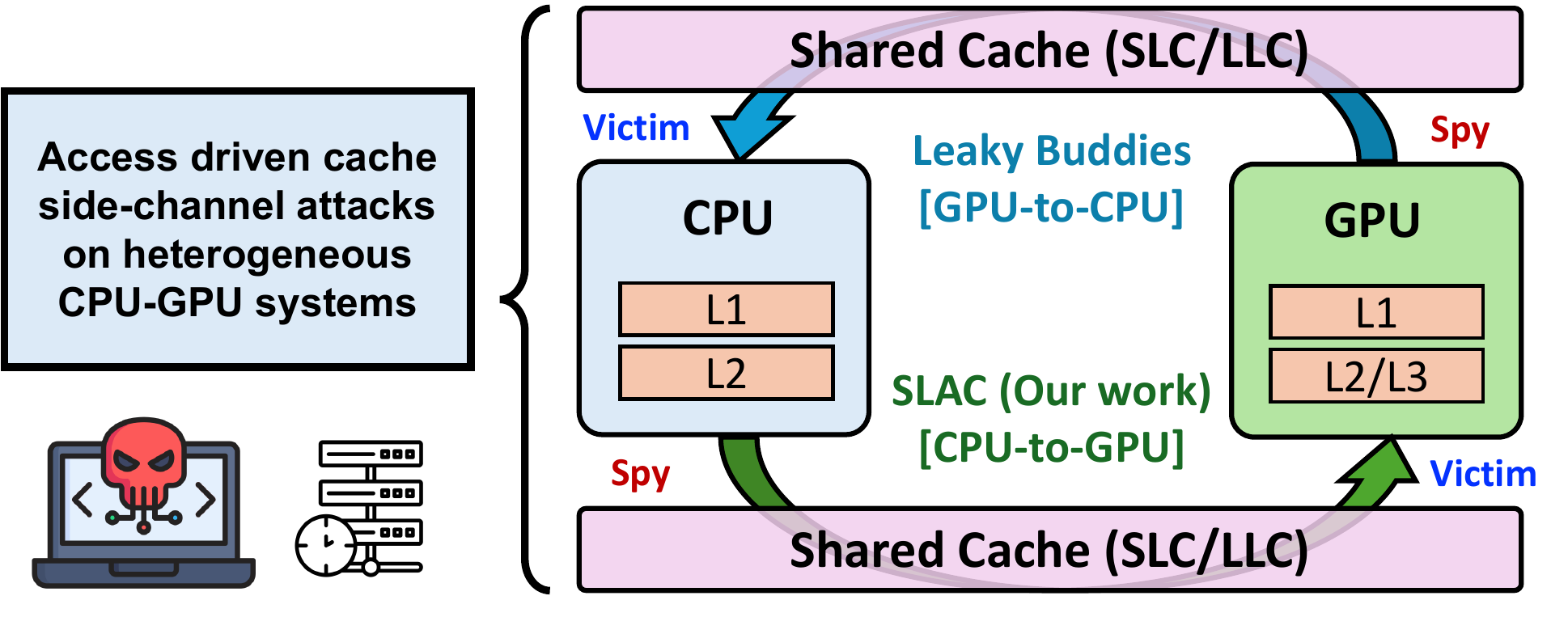}
    \caption{Landscape of access-driven cache side-channel attacks for heterogeneous CPU-GPU architectures.}
    \label{fig:HLV}
\end{figure}

We classify cross CPU-GPU side-channel threats by the respective execution domains of the adversary and the victim. Leaky Buddies~\cite{dutta2021leaky} demonstrates a proof-of-concept GPU-to-CPU side-channel on Intel integrated platforms, in which a GPU-resident adversary monitors the state of the shared Last-Level Cache (LLC) as affected by CPU activities.  The more pressing threat model for sensitive GPU workloads is the opposite CPU-to-GPU direction, in which a CPU-side adversary attacks a GPU victim. Existing attacks in this direction, however, remain limited to coarse-grained cache-occupancy attacks on ARM~\cite{cronin2021exploration} and Apple Silicon~\cite{xu2025exam}, which observe only aggregate cache usage rather than individual cache sets or lines.  Other works also demonstrate bi-directional covert channels through the shared LLC on Intel integrated platforms~\cite{dutta2021leaky, kim2021constructing}. To the best of our knowledge, \textit{no prior work has demonstrated a set-level, access-driven CPU-to-GPU side-channel via a shared cache}. Fig.~\ref{fig:HLV} summarizes the status quo. 

Establishing a fine-grained Prime+Probe side channel from CPU-to-GPU is substantially harder than in earlier results, because the CPU-side adversary must prime the shared cache and reliably observe its state changes at cache-set granularity due to GPU victim activity. This requires a deep understanding of \textit{when}, \textit{where}, and \textit{how} the GPU victim's data accesses affect the shared cache state. Among the three cache inclusion policies~\cite{solihin} that govern shared-cache residency with respect to a private cache, namely exclusive, inclusive, and non-inclusive non-exclusive (NINE), only with the latter two can a victim data access survive the private caches and affect the shared cache state.
Under the exclusive policy, a data access results in cache line residence in the private caches, but
absent from the shared cache, making the shared cache state non-indicative of data accesses.
Under the inclusive policy, every line in the victim's private cache is also always held in the shared cache, while under NINE, the cache line resides in the shared cache conditionally. Only in these two cases can the shared cache reveal victim data access or no-access to a spy process.

We target Apple's M-series SoCs, which have become an attractive platform for privacy-sensitive local AI workloads thanks to their high-performance, energy-efficient GPU acceleration. On Apple Silicon, the SLC is exclusive of the CPU caches and holds a NINE relationship with the GPU caches~\cite{xu2025exam}. Victim information can leak through the SLC via the GPU-side NINE relationship, whereas the CPU-side exclusive relationship governs how the adversary sets up the shared cache and monitors its state changes. A CPU-to-GPU attack is therefore feasible with the specific cache inclusiveness.

The challenge, however, lies in how the adversary primes the shared cache and monitors its state. Because the SLC is exclusive of the CPU caches, a line cannot reside in both the CPU's private caches and the SLC. A CPU adversary has to saturate L2 cache sets first and rely on evictions due to capacity conflicts to spill older cache lines into the SLC. This two-step priming process is fundamentally different from the counterpart used in the prior GPU-to-CPU side-channel on Intel heterogeneous platforms~\cite{dutta2021leaky}. There, the GPU directly primes the LLC via data accesses. However, due to the NINE relationship between the LLC and GPU private caches, primed cache lines may also have a copy in the GPU's private caches. Those copies must be cleared from the GPU's private caches; otherwise, they will interfere with the subsequent probing (access timing measurement) phase, leading to inaccurate observations of victim data activity. In addition to these differences in cache hierarchy and inclusiveness, public knowledge about Apple Silicon's SLC is much less than that of Intel LLC.

For cache attacks on Apple Silicon, the prior work~\cite{xu2025exam} only demonstrates coarse-grained SLC-occupancy attacks, which cannot resolve individual cache lines/sets and therefore cannot recover fine-grained secrets such as ML model inputs or cryptographic keys. Achieving set-level observability further requires architectural knowledge of the Apple SLC that is not publicly available. 

To close this gap, we present \textbf{SLAC}, the first access-driven Prime+Probe SLC side-channel attack on Apple M-series SoCs targeting sensitive GPU workloads. We first develop a reverse-engineering methodology to recover the SLC's set-indexing functions and characterize the interactions between the CPU/GPU private caches and the shared SLC. Building on these findings, we construct \textsc{CPrime+CProbe}, our core CPU-to-GPU side-channel technique, which executes both SLC priming and probing on the CPU and requires no GPU access from the adversary. A covert channel built on \textsc{CPrime+CProbe} achieves a throughput of 62.5~Kbps. We further develop an accelerated extension, \textbf{GPrime}+ \textbf{CProbe}, that primes the SLC from the GPU while retaining the CPU-side probe. GPU parallelism makes priming about  6.4$\times$ faster than \textsc{CPrime}, and the resulting covert channel reaches 400~Kbps. We then demonstrate two end-to-end privacy attacks built on these techniques, namely a graph-structure reconstruction attack on Graph Neural Networks (GNNs) that recovers over 90\% of edges across five datasets, and an input and output retrieval attack on large language models (LLMs) that recovers input keywords with up to 94.8\% accuracy and output tokens with up to 88.9\% accuracy on TinyLlama and GPT-2 Medium.

The contributions of this work include:
\begin{enumerate}[noitemsep,topsep=4pt,leftmargin=13pt]
    \item We reverse-engineer the Apple M1's SLC set-indexing functions and characterize the interactions between private caches and SLC that underpin CPU-observability of GPU memory accesses.
 
    \item We propose the first access-driven CPU-to-GPU side-channel on Apple Silicon, \textbf{CPrime+CProbe},
    and an accelerated variant, \textbf{GPrime+CProbe}, which yields a 6.4$\times$ improvement of the covert-channel throughput.

    \item We demonstrate two end-to-end privacy attacks on GPU workloads: GNN edge recovery that reconstructs over 90\% of edges across five datasets, and LLM input/output recovery that reaches up to 94.8\% and 88.9\% accuracy on TinyLlama and GPT-2 Medium.
    
\end{enumerate}

Our results reveal that cache sharing between CPUs and GPUs in increasingly popular heterogeneous systems poses serious security vulnerabilities and demands careful architectural redesign to ensure the security and privacy of critical GPU workloads. 

The rest of the paper is organized as follows. Section~\ref{sec:back} reviews background and prior relevant work. Section~\ref{sec:reverse} presents our reverse engineering of the Apple M1 SLC. Section~\ref{sec:slc_attack} describes our CPU-to-GPU Prime+Probe side-channel designs and their covert-channel evaluation. Section~\ref{sec:results} demonstrates end-to-end privacy attacks on GPU workloads. Section~\ref{sec:conclusion} concludes the paper with a discussion of future work.

%% file: txt/2.tex
\section{Background and Related Work}\label{sec:back}

This section provides background on cache side-channel attacks and the Apple M-series SoC architecture. We also position our work with respect to prior research on cross CPU-GPU attacks and reverse-engineering of cache set-indexing.

\subsection{Cache Side-Channel Attacks}\label{subsec:CacheSC}

Cache is the most common shared microarchitecture exploited in side-channel attacks, where the  timing difference between cache hit and miss is used to identify cache state change. 
\textit{Cache occupancy attacks} monitor the overall usage or contention level of the shared cache rather than targeting individual cache sets. This side-channel is easy to deploy, robust to noise, and requires only that the spy be familiar with architectural basics, such as the cache inclusion policy.
However, it only provides coarse-grained information leakage, and is typically used for website fingerprinting~\cite{xu2025exam, cronin2021exploration, shusterman2021prime+, shusterman2020website} and data volume-based pattern recognition~\cite{xu2025exam, giner2023scatter, kurth2020netcat}.

The other type, \textit{access-driven side-channel}, offers significantly finer spatial and temporal granularity. It monitors accesses of specific cache sets or lines by the victim. Among them, Prime+Probe is the most widely applicable technique, as it does not require shared memory or special instructions (both are required for Flush+Reload side-channels). The salient steps of setting up Prime+Probe side-channels are:

\begin{itemize}[noitemsep,topsep=5pt,leftmargin=15pt]
  \item \textbf{Prime phase}: The adversary sets the shared cache to a certain state by filling one chosen cache set with its own data. Specifically, an eviction set, which includes a group of adversary data addresses, should be constructed beforehand to be used for priming the cache. The eviction set and the victim secret contend on a cache set, i.e., both map to the cache set through indexing functions over their memory addresses.  
  \item \textbf{Victim execution}: The victim executes its program that may access the secret, therefore causing one replacement on the cache set and evicting one cache line of the adversary.
  \item \textbf{Probe phase}: The attacker re-accesses the eviction set and measures the access latency. A higher latency indicates an eviction by the victim, and hence infers an \texttt{access} of the target victim memory address; while a low latency corresponds to victim \texttt{no-access}.
\end{itemize}

Prime+Probe attacks have been successfully deployed in many scenarios on various CPU architectures~\cite{lipp2016armageddon,genkin2018drive,yan2020cache,purnal2021prime+,shusterman2021prime+,wang2022stealthy}. However, no prior work has realized a Prime+Probe CPU-to-GPU cache side-channel on integrated SoCs.

\subsection{Apple M-series SoCs}\label{subsec:apple}

Apple M-series Systems-on-Chip (SoCs), including the M1, M2, and their subsequent variants, are built with the ARMv8-A Instruction Set Architecture (ISA). Apple M-series SoCs adopt \textbf{Unified Memory Architecture} (UMA), in which the CPU, GPU, and other accelerators access a common pool of physical memory. Apple’s Metal framework, the graphics and compute shader API, provides a unified programming interface for heterogeneous CPU--GPU execution, enabling shared memory to be accessible from both the CPU host and the integrated GPU. This design avoids data copies between host and device memories, thereby improving workload performance.

\begin{figure}[h]
    \centering
\includegraphics[width=0.9\linewidth]{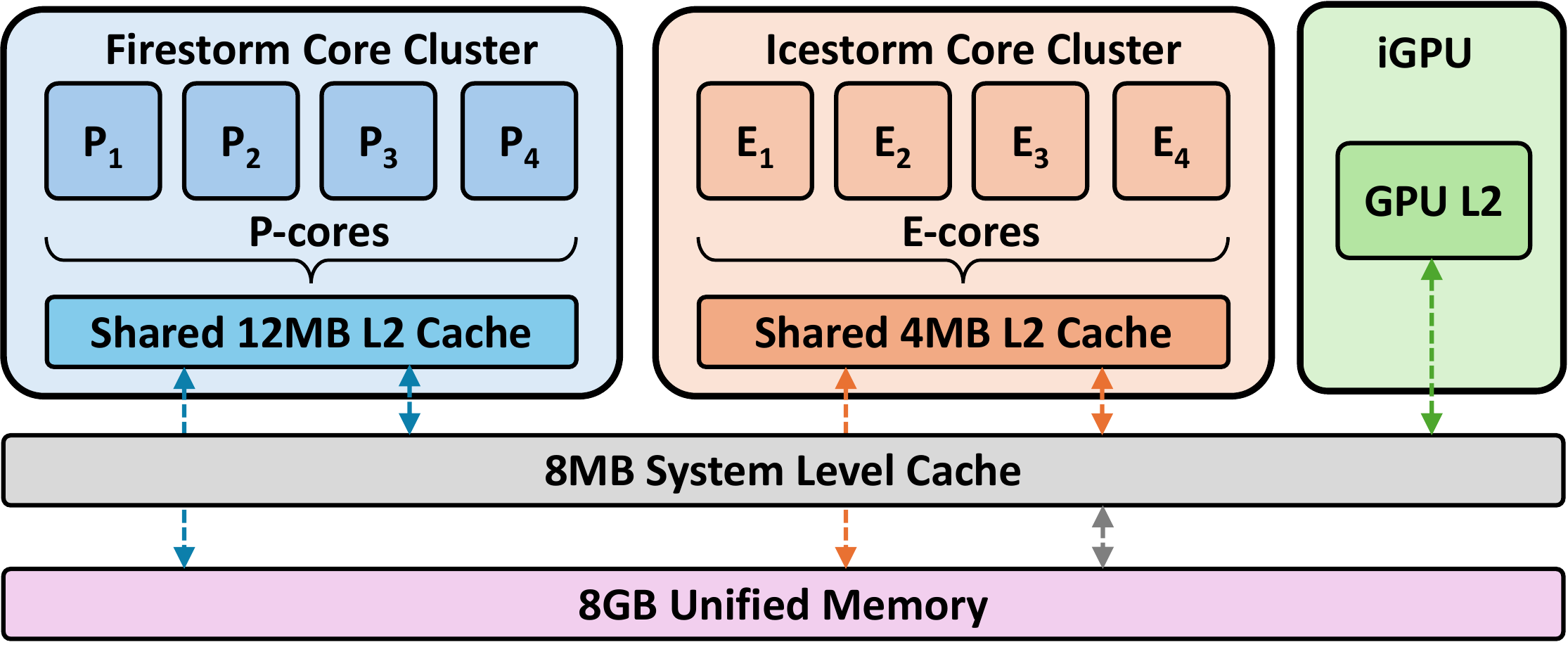}
    \caption{\textit{Cache structure of Apple M1}.}
    \label{fig1}
\end{figure}

Figure~\ref{fig1} depicts the cache hierarchy of Apple M1 and its respective sizes. Specifically, M1 has four high-performance Firestorm cores (p-cores) and four energy-efficient Icestorm cores (e-cores), forming two CPU clusters with core-specific L1 caches and cluster-specific L2 caches. The SLC is shared across the two clusters and the GPU, playing a central role in mediating data access in the UMA design. The cache line size is 128 bytes for all the caches. In practice, macOS schedules active processes onto the Firestorm cluster while reserving Icestorm for low-priority background work. We therefore focus all subsequent experiments on the Firestorm cluster, with CPU being the P-cores and CPU L2 their 12MB L2 cache.

\subsubsection{Prior Reverse Engineering and Cache Attacks on Apple Silicon}
 
Prior work~\cite{xu2025exam} characterized several key properties of the Apple M1 SLC and demonstrated occupancy-based cache attacks on it.

\noindent\textbf{Cache Residency Policy}:
The SLC shows asymmetric cache residency behavior across the CPU and GPU domains. With respect to CPU private caches, the SLC is exclusive: for CPU memory accesses, cache lines are initially kept in the CPU's private caches, and are only ejected to the SLC upon eviction from the CPU L2 cache. In contrast, for GPU memory accesses, corresponding cache lines are directly loaded into the SLC, with a NINE relationship relative to the GPU private caches.
With this insight, an access-driven CPU-to-GPU side-channel is feasible. 

\noindent\textbf{Set Indexing}:
The Apple M1 SLC employs a non-traditional set-indexing scheme that is fundamentally different from CPU local cache indexing. The set index does not use the lowest 13 bits of the physical address, resulting in a sparse, less intuitive mapping between physical addresses and SLC cache sets and complicating eviction-set construction. 
There is no prior public knowledge of Apple Silicon SLC set indexing functions.
 
\noindent\textbf{Occupancy-based Cache Attacks}:
The prior work~\cite{xu2025exam} only demonstrated cache-occupancy attacks on Apple M1, including website fingerprinting and screen capture via aggregate SLC usage. Fine-grained, access-driven CPU-to-GPU leakage on Apple Silicon, however, remains unexplored and is the focus of this work.

\subsection{Related Work on Cross CPU-GPU Platforms}\label{subsec:related}
 
Cross CPU-GPU Prime+Probe cache side-channels on integrated SoCs have so far been studied only on Intel platforms, in Leaky Buddies~\cite{dutta2021leaky} with proof-of-concept preliminary results.
Cache set-indexing functions on Intel platforms have been extensively reverse-engineered in prior work~\cite{gruss2016rowhammer, hund2013practical, vila2019theory, morgan2025slice+, yu2023synchronization}, which showed that Intel's LLC set indexing uses XOR hash functions over 6 bits or more of the physical address. Recovering these functions has enabled practical cache attacks on X86 systems. Beyond Intel, however, set-indexing for other heterogeneous CPU-GPU systems remains far less understood, particularly on platforms where the shared last-level cache is not inclusive of the CPU private caches. Recent work on AMD processors~\cite{wang2025zenleak}
only demonstrated cross-core side-channel leakage. On Apple's M-series SoCs, prior work~\cite{xu2025exam} only coarsely characterized the SLC indexing without resolving the concrete hash functions. We close this gap by reverse-engineering the SLC's set-indexing functions in Section~\ref{sec:reverse} and then building our side-channel techniques in Section~\ref{sec:slc_attack}.

%% file: txt/3.tex
\section{Reverse Engineering Apple M1's SLC Set Indexing Functions}\label{sec:reverse}

This section presents our reverse engineering of the Apple M1's  SLC set indexing functions,
which are a prerequisite for constructing the access-driven SLC side-channel, 
as targeting a specific cache-set depends on knowing how physical addresses map to sets. This reverse engineering process, however, is non-trivial for two reasons.

\begin{figure*}[h]
    \centering
    \begin{subfigure}{0.33\linewidth}
        \centering
        \includegraphics[width=0.8\linewidth]{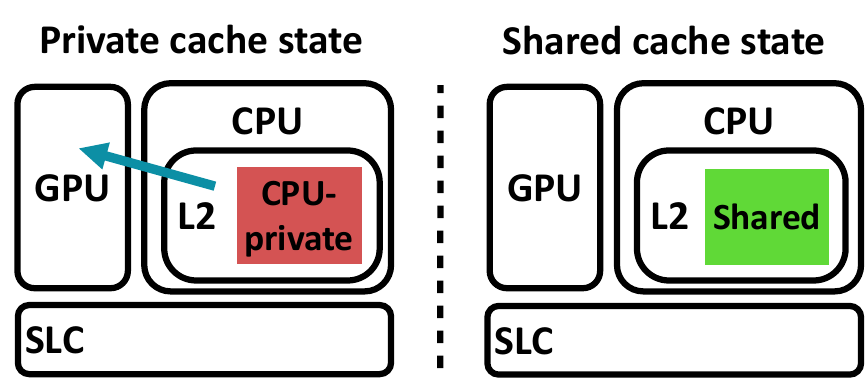}
        \caption{\ding{192}: CPU-private (L2-resident) to shared.}
        \label{fig:shared-data-case-a}
    \end{subfigure}
    \begin{subfigure}{0.33\linewidth}
        \centering
        \includegraphics[width=0.8\linewidth]{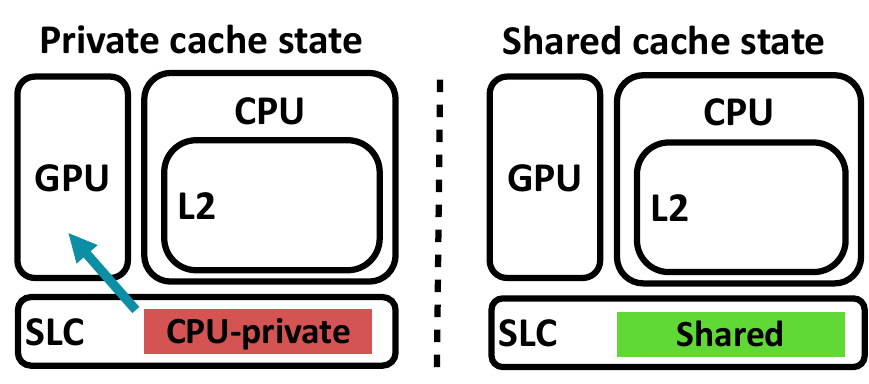}
        \caption{\ding{193}: CPU-private (SLC-resident) to shared.}
        \label{fig:shared-data-case-b}
    \end{subfigure}
    \begin{subfigure}{0.33\linewidth}
        \centering
        \includegraphics[width=0.8\linewidth]{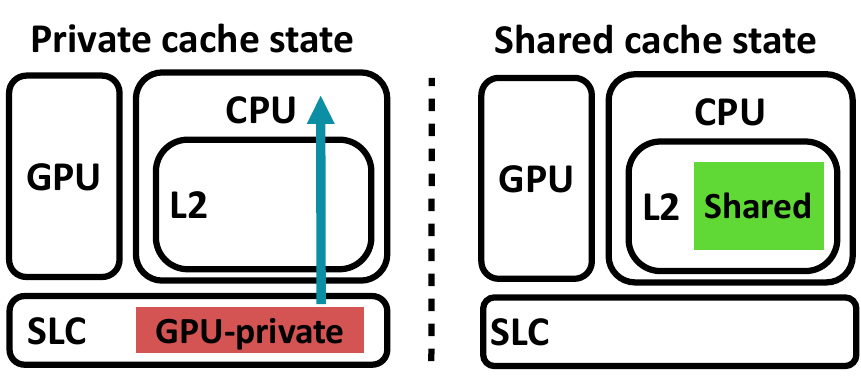}
        \caption{\ding{194}: GPU-accessed to shared. }
        \label{fig:shared-data-case-c}
    \end{subfigure}
    \caption{Three shared-data transition scenarios, showing an example cache line's initial private location (red), the data access that triggers sharing (arrow), and the resulting shared location (green).}
    \label{fig:shared-data-cases}
\end{figure*}

\noindent\textbf{Challenge 1: CPU-exclusive SLC residency.}
The SLC is exclusive with respect to the CPU private caches,
which means that CPU-side data accesses cannot reliably populate or locate SLC cache lines. Standard CPU-based eviction-set construction, therefore, does not apply, and set congruency cannot be inferred from the timing of CPU data access.
 
\noindent\textbf{Challenge 2: Fully hashed set indexing.}
As noted in Section~\ref{subsec:apple}, the SLC does not draw its set index from contiguous low-order physical address bits directly. 
Each cache index bit is derived from a hash function over multiple address bits.
This mapping expands the candidate space for eviction-set construction and makes traditional iterative pruning unstable under realistic measurement noise.
 
We overcome these two challenges in our reverse engineering. Section~\ref{subsec:challenge1} examines how \emph{shared data} is handled between the CPU and GPU, and shows that GPU data accesses can place cache lines into the SLC while the CPU observes their residency, resolving Challenge~1. Building on CPU-side timing measurement, Section~\ref{set_index} introduces an eviction-set construction method that replaces noise-sensitive iterative pruning with collision-profile clustering, thereby addressing Challenge~2.

\subsection{SLC Behavior in CPU-GPU Data Sharing}\label{subsec:challenge1}

We closely track how CPU--GPU \emph{shared data} moves through the system cache hierarchy, between the SLC and CPU private caches.
We classify shared data into three representative scenarios based on its initial residency state before sharing:

\begin{itemize}[noitemsep,topsep=5pt,leftmargin=10pt]
    \item[\ding{192}] \textbf{CPU-private (L2-resident) to shared.}
    Data initially resides in the CPU's private L2 cache and is not present in the SLC due to the L2 cache's exclusivity. The GPU subsequently accesses this data block.

    \item[\ding{193}] \textbf{CPU-private (SLC-resident) to shared.}
    Data was previously evicted from the CPU's L2 cache into the SLC. The GPU then accesses this SLC-resident data.

    \item[\ding{194}] \textbf{GPU-accessed to shared.}
    Data is initially accessed by the GPU and resides in the SLC, and the CPU accesses it thereafter.
\end{itemize}

These three scenarios capture all possible transitions of cached data from a private state to a shared state. For each scenario, we use subsequent CPU-side access and measure its latency to distinguish among CPU local cache hits (L1 or L2), SLC hits, and SLC misses, thereby tracking the precise residence at specific cache levels.
Figure~\ref{fig:shared-data-cases} summarizes the results for the three scenarios. Note that we focus on the SLC and CPU cache states rather than the GPU cache state, due to the lack of reliable GPU-side timing measurements on Apple Silicon~\cite{zhang2024invalidate+, jiang2016complete}.

\begin{itemize}[noitemsep,topsep=5pt,leftmargin=15pt]
    \item[\ding{192}] After the GPU accesses a line that resides in the CPU private caches, subsequent CPU accesses continue to hit in local caches, and no new copy of the line appears in the SLC (Appendix~\ref{app:slc-copy}). This reveals part of the cache coherency: the GPU reads the line directly from the CPU L2 cache, changing the CPU-resident copy to a shared state ~\cite{apple2024scalable, apple2021intercluster}. 
    \item[\ding{193}] After the GPU accesses data originally residing in the SLC, the CPU’s first subsequent access results in an SLC hit, and later accesses yield CPU local cache hits, indicating that GPU access does not invalidate the SLC copy and only changes it to a shared state.
    \item[\ding{194}] When the CPU first accesses data previously accessed only by the GPU, the access results in an SLC hit. Subsequent CPU accesses to the same data are then consistently served by the local
    caches. 
\end{itemize}

These observations identify two constraints on reliable SLC filling and observation. Scenario \ding{192} shows that a GPU access does not evict a line already held in the CPU L2 to SLC, so a GPU read alone cannot always fill an SLC cache line. 
Scenarios \ding{193} and \ding{194} show that the CPU detects SLC residency only on its first access and that all subsequent accesses find the line in the CPU local cache. 
Therefore, an SLC-residency probe must ensure that the target line is not already cached in the CPU's local cache hierarchy before each measurement.

\if false 
We design a shared-buffer measurement framework. 

\begin{enumerate}[noitemsep,topsep=4pt,leftmargin=15pt]
    \item \textbf{Clear CPU-local residency.} We remove the target lines from CPU L2, either by accessing a large unrelated buffer from the CPU or GPU, or by having the GPU write to the shared buffer.
    \item \textbf{Fill the SLC from the GPU.} The GPU sweeps the entire target buffer, which loads every line into the SLC.
    \item \textbf{Probe from the CPU.} We access the same buffer from the CPU and measure per-line latency to determine which lines reside in the SLC.
\end{enumerate}

This three-step procedure lets us control and observe SLC residency through shared memory. Building on it, Section~\ref{set_index} turns set-congruency inference into an eviction-set construction method that addresses Challenge~2.

\fi 

\subsection{Eviction Set Construction}\label{set_index}

Recovering the set-index function requires eviction sets, that is, groups of addresses that all map to the same cache set. Most prior methods construct eviction sets by iterative pruning. They start from a large candidate pool of memory addresses, repeatedly test whether subsets can evict a reference line, and shrink the pool until only a minimal congruent set remains~\cite{gruss2016rowhammer,hund2013practical,vila2019theory,morgan2025slice+,yu2023synchronization}. The size of this initial pool scales with cache associativity multiplied by $2^{u}$, where $u$ is the number of set-index bits generated by hashing. In the prior settings for these methods, only a small subset of the index bits is hashed while the remaining bits are drawn directly from low-order physical address bits, so 
$u$ is usually fewer than eight, and the candidate pool stays within a few thousand lines.

The Apple M1 SLC uses a substantially different set-indexing mechanism. Prior analysis~\cite{xu2025exam} indicates that none of the set-index bits are drawn directly from the physical address. Section~\ref{subsec:recover-slc} shows that all 12 SLC set-index bits are computed by hash functions over multiple address bits, causing the initial candidate pool to expand to tens of thousands of cache lines. Such a large search space makes eviction-set construction highly sensitive to background noise and prevents the standard iterative pruning process from converging reliably.

Instead, we construct eviction sets using \emph{collision-profile clustering}, which replaces single-line eviction testing with a richer per-address signature. The procedure has three steps. We first
fill the SLC from the GPU side (details given in Section~\ref{sec:gprime}) with a filler buffer $\mathcal{B}_{\mathrm{fill}}$ containing many cache-line-sized memory blocks. We then access a single test cache line from the GPU via another buffer $\mathcal{B}_{\mathrm{test}}$, which contends for SLC residency with filled cache lines. After this access, we probe all lines in $\mathcal{B}_{\mathrm{fill}}$ from the CPU and record which filler lines were evicted. The resulting eviction pattern forms the \emph{collision profile} of that test line. We repeat this measurement for every test line in $\mathcal{B}_{\mathrm{test}}$, producing one collision profile per line.

Two test lines that map to the same SLC set will evict the same filler lines and therefore produce highly similar collision profiles. We exploit this property by clustering test lines based on profile similarity. Each cluster corresponds to one SLC set, and the addresses within a cluster form an eviction set. Because each clustering decision aggregates information from many filler-line observations rather than a single hit-or-miss test, this approach is substantially more tolerant to background noise, scheduling jitter, and transient cache-state fluctuations than iterative pruning. The resulting clusters provide reliable inputs for recovering the explicit set-index hash functions.

\subsection{Recovering SLC Set Indexing Functions}\label{subsec:recover-slc}

We apply the collision-profile clustering method to a test region $\mathcal{B}_{\mathrm{test}}$ whose size modestly exceeds the SLC capacity, to ensure that the entire SLC is filled even when the order of filling is non-uniform.
We set $|\mathcal{B}_{\mathrm{test}}|$ to $1.2\times$ the SLC size in our experiments. The clustering yields 4,096 clusters, indicating 4,096 SLC sets and therefore a 12-bit set index. We obtain the physical addresses of the cache lines in each cluster through the macOS I/O Kit (\texttt{IOMemoryDescriptor}) and then recover a functionally equivalent set of indexing functions as follows:

\begin{enumerate}[noitemsep,topsep=4pt,leftmargin=16pt]
    \item \textbf{Identify candidate address bits.} On our 8GB Apple M1 platform, the lower 33 bits compose the effective physical byte address. With 128-byte cache lines, the lowest 7 bits are offset bits, leaving 26 candidate bits contributing to the 12-bit SLC set index.

    \item \textbf{Build candidate function library.} We model each index bit as an XOR over a subset of the 26 candidate bits, yielding a library of $2^{26}$ candidate bit functions.

    \item \textbf{Filter valid bit functions.} A valid index-bit function must output the same value for all addresses within a cluster, and across 4,096 clusters, it should split evenly (half 0, half 1). This reduces the candidates to 4,095 valid bit functions.

    \item \textbf{Select linearly independent functions.} We select 12 linearly independent valid bit functions that jointly map the 4,096 clusters to distinct 12-bit indices.
\end{enumerate}

\begin{figure}[h]
    \centering
    \includegraphics[width=0.9\linewidth]{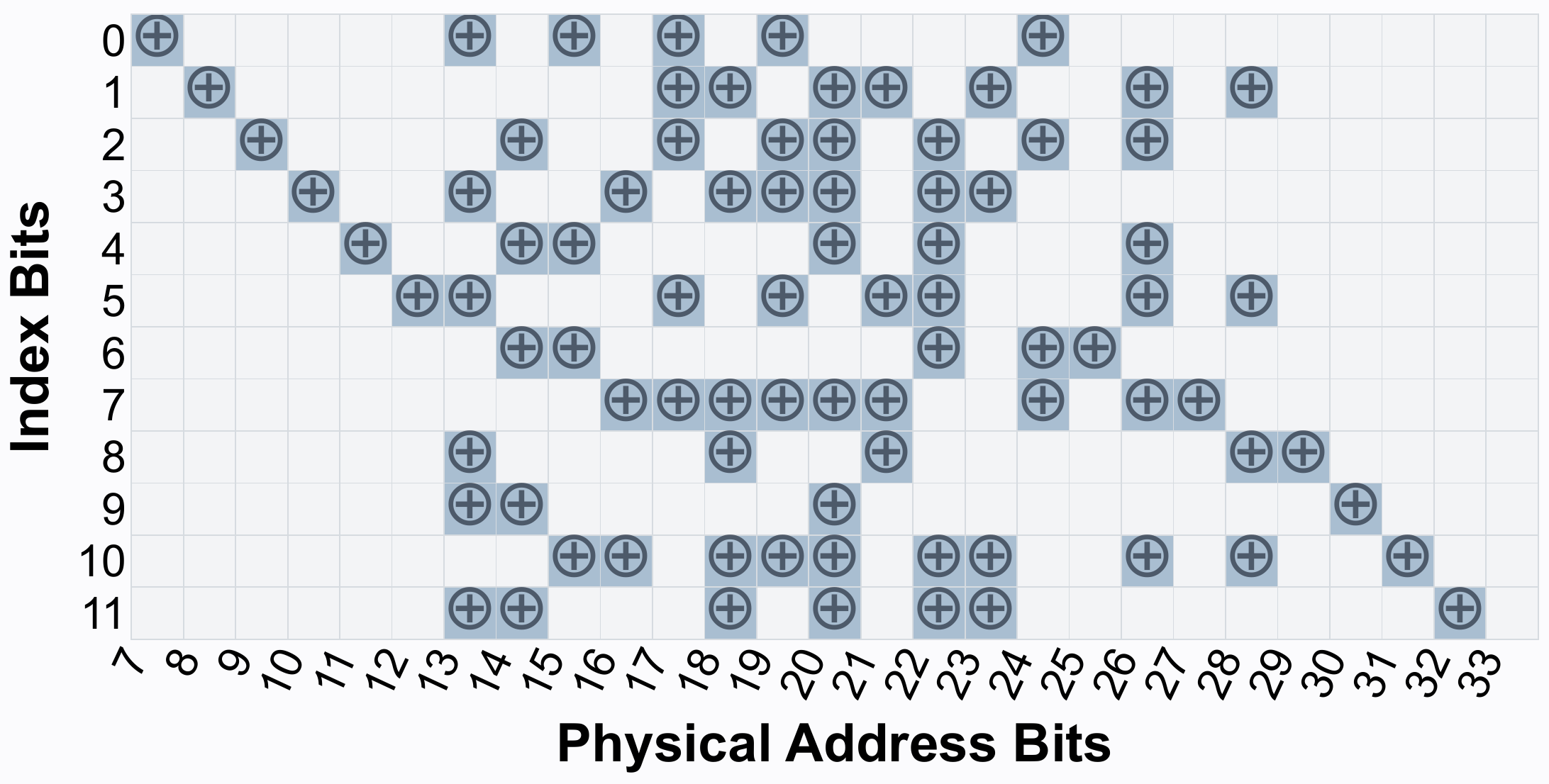}
    \caption{\textit{One set of 12 hashing functions for system-level cache (SLC) set index bits.}}
    \label{fig:hashheatmap}
\end{figure}

Any such set of 12 bit functions is functionally equivalent. The numeric set labels produced by a selected set may differ from the hardware-internal labels, but the recovered mapping deterministically partitions physical addresses into 4,096 distinct SLC sets. Figure~\ref{fig:hashheatmap} shows one set of 12 index bit functions over physical address bits 7--32.

\noindent\textbf{Replacement policy.}
We use a similar process to characterize the SLC replacement policy. 
We construct a larger eviction set for a selected SLC cache set (twice the cache associativity).  
We execute GPU kernels that access these lines in controlled orders. CPU probing after each kernel reveals which lines have been evicted. Across configurations, eviction is consistently access-order-dependent: lines accessed earlier are replaced before those accessed more recently, which is consistent with an LRU replacement policy~\cite{xiong2020leaking,lee1999existence}.
 
With the set-indexing function and replacement policy now characterized, we can construct eviction sets that target any SLC set we choose and can predict the eviction order within each set. These two ingredients underpin the CPU-to-GPU side-channel techniques developed in the next section.

%% file: txt/4.tex
\section{CPU-to-GPU Prime+Probe SLC Side-channel on Apple M1}\label{sec:slc_attack}

Building on our reverse-engineering results, we develop an access-driven cache side-channel technique over the Apple M1 System-Level Cache (SLC) with integrated CPU-GPU. The mechanism adapts the classic \textit{Prime+Probe} paradigm to monitor fine-grained SLC set activity induced by GPU workloads.

\subsection{Overview of the Side-channel}\label{sec:overview}

Our threat model includes a sensitive victim workload on the GPU and an adversary on the CPU to carry out SLC priming and probing. The probing phase is always performed on the CPU (\textsc{CProbe}, Section~\ref{sec:cprobe}) to distinguish SLC hits and misses through accurate access-latency measurements, because Apple GPUs lack reliable fine-grained timing mechanisms~\cite{zhang2024invalidate+, jiang2016complete}.
For the priming phase, which populates target SLC sets with attacker-controlled eviction sets, we introduce two alternative designs that differ in efficiency and in the capabilities required of the attacker. The CPU-based strategy \textsc{CPrime} (Section~\ref{sec:cprime}) executes entirely on the CPU and does not require the adversary to have GPU access, which makes it easier to deploy under restrictive threat models. When the adversary can also execute on the GPU, the GPU-based strategy \textsc{GPrime} (Section~\ref{sec:gprime}) exploits GPU parallelism to substantially accelerate priming. With the two different techniques, we build high-bandwidth covert channels and quantify their throughputs (Section~\ref{sec:covert}).

\begin{figure*}[t]
    \centering
    \includegraphics[width=0.95\linewidth]{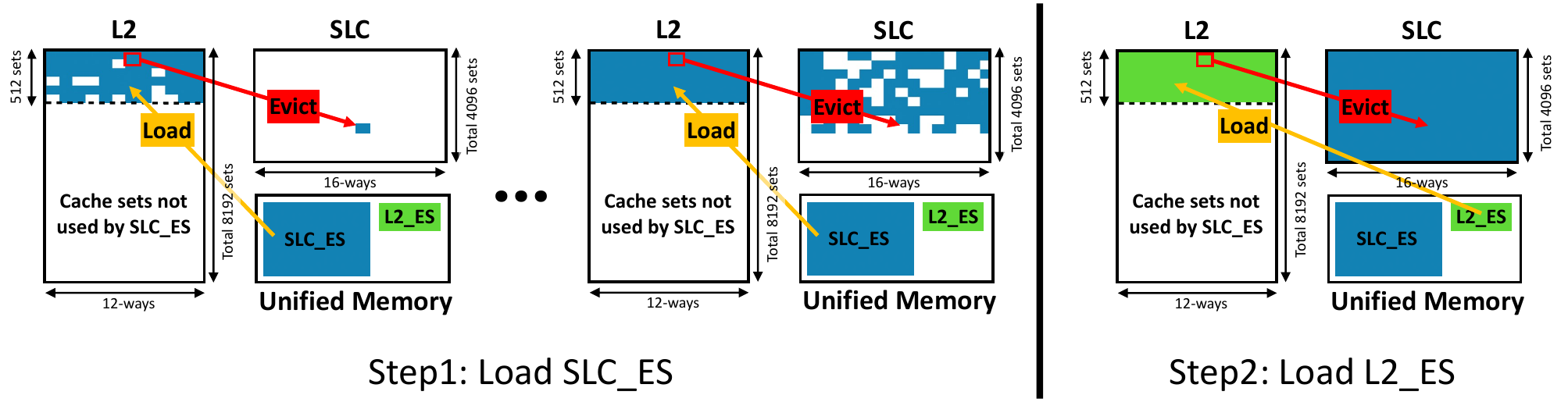}
    \caption{\textsc{CPrime} procedure: load SLC\_ES first (cache lines only partially in SLC, the rest reside in L2), then load L2\_ES.}
    \label{fig:Cprime}
\end{figure*}

\subsection{Prime with CPU (\textsc{CPrime})}\label{sec:cprime}

On Apple M1, the SLC is exclusive with respect to the CPU private caches, so a cache line resides in either the L2 cache or the SLC, but never in both simultaneously. Also, the SLC acts as a backup buffer: when the CPU loads data from memory, it always goes to the L2 cache. Only when an L2 cache set is full, an older cache line is evicted to the SLC due to capacity constraints.
To prime the SLC from the CPU, the attacker must load more data than the SLC capacity, using fresher data to evict older data from L2 to fill the SLC.  Because the SLC and L2 use different indexing functions, memory blocks chosen to target specific SLC sets are generally mapped to multiple L2 sets, increasing the L2 footprint and consequently slowing the priming process. Considering the exclusiveness, \textsc{CPrime} requires two coordinated data eviction sets: one eventually fills the SLC, and another is the evictor - fresher data that is loaded into L2 while evicting older cache lines to the SLC.

\noindent\textbf{\textsc{CPrime} setup.}
We now describe these two eviction sets in detail. The first set, \textit{SLC\_ES}, aims to fill the entire SLC. With the Apple M1 SLC set associativity of 16 and 4,096 cache sets, \textnormal{SLC\_ES} contains $4{,}096 \times 16 = 65{,}536$ memory addresses in total (each corresponds to one cache-line size memory block). The second set of memory addresses, \textit{L2\_ES}, tracks all CPU L2 cache sets affected by memory blocks in \textnormal{SLC\_ES} and aims to fill those sets. Given that the Apple M1 CPU L2 cache set associativity is 12, \textnormal{L2\_ES} contains $n_{L_2}\times12$ memory addresses (cache-line size memory blocks), where $n_{L_2}$ is the number of affected L2 cache sets.

\noindent\textbf{Priming procedure.}
\textsc{CPrime} proceeds in two steps, as shown in Figure~\ref{fig:Cprime}. In the first step, the attacker issues one load to each of the 65,536 cache-line addresses in \textnormal{SLC_ES}, accessing them in sequence so that every block enters L2. As loading continues, some  L2 cache sets
reach their 12-way capacity, causing the earliest-loaded blocks at those sets to be evicted from L2 into the SLC. At the end of this step,  early-loaded blocks of \textnormal{SLC_ES} reside in the SLC, while  recent loads reside in the L2 at affected  sets. 
In the second step, the attacker loads \textnormal{L2_ES}, which contains 12 addresses for each affected L2 set, which fill the set entirely and push out the surviving \textnormal{SLC_ES} data blocks into the SLC. At the end of this step, \textnormal{SLC_ES} successfully occupies the SLC in full, completing the priming, while \textnormal{L2_ES} occupies the corresponding L2 sets.

\noindent\textbf{Design optimization.} The only variable of the \textsc{CPrime} process is
the number of L2 sets affected by \textnormal{SLC\_ES} ($n_{L_2}$).  We carefully choose \textnormal{SLC\_ES} so as to reduce $n_{L_2}$, lowering the priming time.

\if false 

\item \textit{How are the eviction sets constructed without knowledge of physical addresses?}
\fi 
A natural approach for constructing an eviction set is to apply the set-indexing functions recovered in Section~\ref{set_index} directly to physical addresses. However, macOS does not expose physical addresses to user-space processes. We therefore build both eviction sets through timing-based primitives and only use the indexing functions to validate them. \textnormal{L2\_ES} is constructed by the traditional iterative pruning method~\cite{gruss2016rowhammer,hund2013practical,vila2019theory,morgan2025slice+}.
~ This procedure is reliable for L2 because its index uses only a handful of address bits, thereby keeping the candidate pool small. \textnormal{SLC\_ES} is constructed by the collision-profile clustering method presented in Section~\ref{set_index}, which identifies set-congruent addresses from the collisions they induce between a test region $\mathcal{B}_\mathrm{test}$ and a filler region $\mathcal{B}_\mathrm{fill}$.

The L2 cache derives most of its set index bits directly from physical address bits 7--13 and 16--19, while the remaining two index bits are generated through hashing over higher physical address bits\cite{yu2023synchronization}. In contrast, all SLC set-index bits are computed through hash functions over multiple physical address bits. Because the SLC hash distributes addresses nearly uniformly, a naively chosen \textnormal{SLC\_ES} may spread across all 8,192 L2 sets, i.e., yielding a large  $n_{L_2}$ and resulting in  \textnormal{L2\_ES} with $8{,}192 \times 12 \approx 98\text{K}$ addresses. Loading such a large eviction set
for priming would be prohibitively slow. We reduce this cost by constraining address bits 7--10 to zero when constructing \textnormal{SLC\_ES}, as they fall within the page offset and can be controlled through virtual addresses alone.
Fixing these four bits restricts \textnormal{SLC\_ES} to $1/16$ of the L2 cache sets, lowering $n_{L_2}$ from 8,192 to 512.

Fixing additional address bits would further shrink $n_{L_2}$, but at the cost of a larger page footprint for the eviction set. The page size on macOS is 16\,KB, holding 128 cache-line blocks (every 128 bytes) with their page offsets specified by the address bits 7--13. Fixing each bit halves the usable memory blocks per page. For example, when bits 7--10 are fixed to zero, $2^{7-4} = 8$ blocks per page are usable for \textnormal{SLC\_ES} , and therefore the eviction set spans $65{,}536 / 8 = 8{,}192$ pages. Fixing an additional bit would increase the page footprint to 16,384 pages. We found that, under this larger page footprint, macOS frequently remaps pages during the attack, invalidating the constructed eviction sets and preventing reliable priming. Such remapping rarely occurs when four bits are fixed. Therefore, we fix this at 4 bits, which reduces the L2 cache set footprint by 16$\times$ while keeping the page count at half the remapping threshold.

\subsection{Prime with GPU (\textsc{GPrime})}\label{sec:gprime}

The attacker can also prime the SLC using a GPU to leverage the massive parallelism, i.e.,  \textbf{GPrime}. There is another advantage due to asymmetric cache inclusiveness: SLC is exclusive to CPU caches but not exclusive to GPU caches. GPU memory accesses can directly populate the SLC, and therefore, one eviction set \textnormal{SLC\_ES} is sufficient for priming the SLC.
However, a complication arising from cache coherence between the CPU and GPU needs to be addressed.

\noindent\textbf{Invalidating shared data in CPU L2 from GPU.}
Based on findings presented in Section~\ref{subsec:challenge1} (Case 1), if a shared cache-line memory block already has a copy in CPU private caches, GPU read can make a copy in GPU caches through cache coherency without going through the SLC. Therefore, directly reading \textnormal{SLC_ES} from the GPU may fail to realize SLC priming when there are cache-line copies in CPU L2. We therefore introduce a GPU-write step to invalidate these stale CPU private-cache copies. However, GPU write-only cannot prime the SLC. Our experiments show that the GPU write operations do not allocate cache lines into the SLC.
 
This no-write-allocate behavior with SLC is consistent with common GPU cache designs, where write traffic is often bypassed to avoid cache pollution~\cite{agarwal2016selective,singh2013cache}.
We further find that even a read-after-write sequence within the same kernel does not reliably bring the data into the SLC. A likely explanation is that the GPU optimizes such producer-consumer data accesses using on-chip tile memory~\cite{apple_tile_based_rendering}, so the subsequent read can be serviced directly from the to-be-written value, rather than by first committing the write to memory and then reloading the data through the SLC path. Based on these findings, we design \textsc{GPrime} with two GPU kernels: the first kernel writes to all \textnormal{SLC_ES} memory blocks to invalidate any stale CPU private-cache copies or SLC copies, and the second kernel reads the same \textnormal{SLC_ES} addresses to load these memory blocks into the SLC cache lines.

\noindent\textbf{Facilitating parallel memory accessing.}
With the massive parallelism offered by the single-instruction-multiple-thread (SIMT) programming model of GPUs, both GPU-write and GPU-read can be accelerated by processing many data block addresses concurrently. We maintain an auxiliary \textit{index\_sets} structure that stores indices of \textnormal{SLC\_ES} addresses for parallel dispatch across GPU threads. This organization is necessary because the method for accessing an eviction set to prime a cache from the CPU does not work for the GPU.  The common eviction-set access method is pointer-chasing~\cite{xiong2020leaking,hund2013practical,vila2019theory,morgan2025slice+}, in which a memory block contains the next address to access, and accesses are sequentially data-dependent, which do not fit thread-level GPU parallelism. To avoid \textit{index_sets} itself entering the SLC and introducing additional noise, the attacker performs a single pass of CPU reads of \textit{index_sets} to bring them into the CPU L2 before launching the GPU kernels. During \textsc{GPrime} kernel execution, the GPU accesses this set via cache coherency, allowing \textit{index_sets} to remain in CPU L2 throughout the attack without entering the SLC.

Although \textsc{GPrime} involves both a write kernel and a read kernel, it remains significantly more efficient than \textsc{CPrime}. In both kernels, one GPU thread is assigned to each address in \textnormal{SLC_ES}, resulting in 65,536 threads per kernel. We set the \texttt{threadgroup} size to 1,024 threads, the maximum supported by the Metal API on Apple M1.
Under this configuration, the kernel threads are scheduled in batches of \texttt{threadgroups} and collectively exploit the GPU’s available execution resources at a high degree of parallelism. As a result, \textsc{GPrime} completes more than an order of magnitude faster than \textsc{CPrime}, reducing the total priming time from 7 ms to 0.7 ms. This speedup is attributed to both the massive thread parallelism and the elimination of  \textnormal{L2\_ES} for \textsc{GPrime}.

\subsection{Probe (\textsc{CProbe})}\label{sec:cprobe}
Once the SLC is primed and the victim GPU kernel has finished execution, the attacker must identify the SLC sets accessed by the victim. \textsc{CProbe} performs this step from the CPU side by re-accessing every memory block in \textnormal{SLC\_ES} and timing each access. 
As our priming process ensures no copies in CPU local caches,
each \textsc{CProbe} access can resolve to either an SLC hit or an SLC miss. An SLC-hit access indicates that the primed line is still resident in the SLC, so the victim did not access any of its own data that maps to this set. An SLC-miss access indicates that the primed line has been evicted, most likely because the victim accessed data mapping to this set, and sometimes because of background system activity.

\noindent\textbf{Fine-grained occupancy measurement.}
Unlike conventional \textit{Prime+Probe} attacks that record only a binary access-or-no-access outcome per cache set, \textsc{CProbe} counts how many of the 16 primed lines in each SLC set yield high-latency accesses.
This per-set integer reveals not only which SLC sets the victim accessed but also how heavily it used each of them, which carries substantially more information than a binary access signal. The final output of \textsc{CProbe} is therefore an integer trace of length 4{,}096 (one entry per SLC set), with each element taking a value in $\{0, 1, \dots, 16\}$ that represents the number of SLC misses observed for the corresponding set.

\subsection{Covert Channels}\label{sec:covert}

Before applying our side-channel techniques against real GPU workloads in Section~\ref{sec:results}, we first measure their raw leakage capacity through covert channels. A covert channel is a controlled setting in which the sender and receiver collude to communicate via a microarchitectural medium, in our case, the SLC. This setting removes the confounding effects of a real workload, allowing us to cleanly quantify the throughput, decoding accuracy, and noise tolerance for the two side-channel variants \textsc{CPrime}+\textsc{CProbe} and \textsc{GPrime}+\textsc{CProbe}.

\noindent\textbf{Communication protocol.}
A GPU kernel acts as the sender, and a CPU process acts as the receiver. In each communication cycle, the receiver first primes a designated SLC set, the sender then modulates the state of that set, and the receiver finally probes the set to recover the transmitted bit. To transmit a `1', the sender fills all 16 cache lines of the target set, resulting in zero SLC hits at probe time.
To transmit a `0', the sender remains idle on that set, so every primed line survives and all 16 produce SLC hits. This full-set modulation maximizes the separation between the two symbols and makes the channel robust to background noise such as OS-induced cache traffic.
Under the SLC's LRU replacement policy, the last-primed lines are the most recently used in the set, and any partial eviction from background activity displaces older lines first. If even the last-primed lines are missing at probe time, the set has almost certainly been filled by the sender. The receiver therefore probes only the last five primed addresses, which keeps the probe phase short while preserving decoding reliability.

\noindent\textbf{Encoding granularity and noise resilience.}
In the protocol above, each cache set encodes one bit. The viable encoding granularity, however, depends on the priming method and the noise it introduces. With \textsc{GPrime}, GPU-based priming populates the SLC in a single uniform pass and produces a clean, low-noise baseline, so one cache set per bit already achieves above 99\% bit-level decoding accuracy. \textsc{GPrime}+\textsc{CProbe} therefore uses this one-set-per-bit encoding. \textsc{CPrime}, in contrast, relies on a two-stage priming procedure that involves CPU L2 evictions and is exposed to interference from the OS and CPU private caches. Single-set decoding accuracy falls to 88\% under these conditions. To restore reliable transmission, \textsc{CPrime}+\textsc{CProbe} assigns four cache sets per bit and has the receiver sum the probe hits across the four sets before decoding. This aggregation improves the decoding accuracy to be above 99\%, at the cost of a 4$\times$ reduction in bit density per communication cycle.

\noindent\textbf{Synchronization and throughput.}
The sender and receiver are synchronized using a millisecond-level timer to align the three phases of priming, transmission, and probing (decoding) within each communication cycle.
The two side-channel mechanisms differ in the priming time and transmission density.
The priming times reported in Section~\ref{sec:gprime} (about 7\,ms for \textsc{CPrime} and 0.7\,ms for \textsc{GPrime}) are run-to-run averages, and individual runs fluctuate around these values. \textsc{CProbe} likewise takes about 5\,ms on average per cycle. To ensure each phase completes before the next begins, even in the worst case, we provision each slot slightly above the average execution time: an 8\,ms slot for \textsc{CPrime}, a 2\,ms slot for \textsc{GPrime}, and a 6\,ms slot for \textsc{CProbe}.
In one \textsc{GPrime}+\textsc{CProbe} cycle, 4{,}096 bits are transmitted, while only 1{,}024 bits are transmitted per \textsc{CPrime}+\textsc{CProbe} cycle. Taking into account both the cycle time and the encoding granularity, \textsc{CPrime}+\textsc{CProbe} achieves a throughput of 62.5\,Kbps, while \textsc{GPrime}+\textsc{CProbe} reaches 400\,Kbps.

\if false 
\begin{figure}[t]
    \centering
    \begin{subfigure}{\linewidth}
        \centering
        \includegraphics[width=0.8\linewidth]{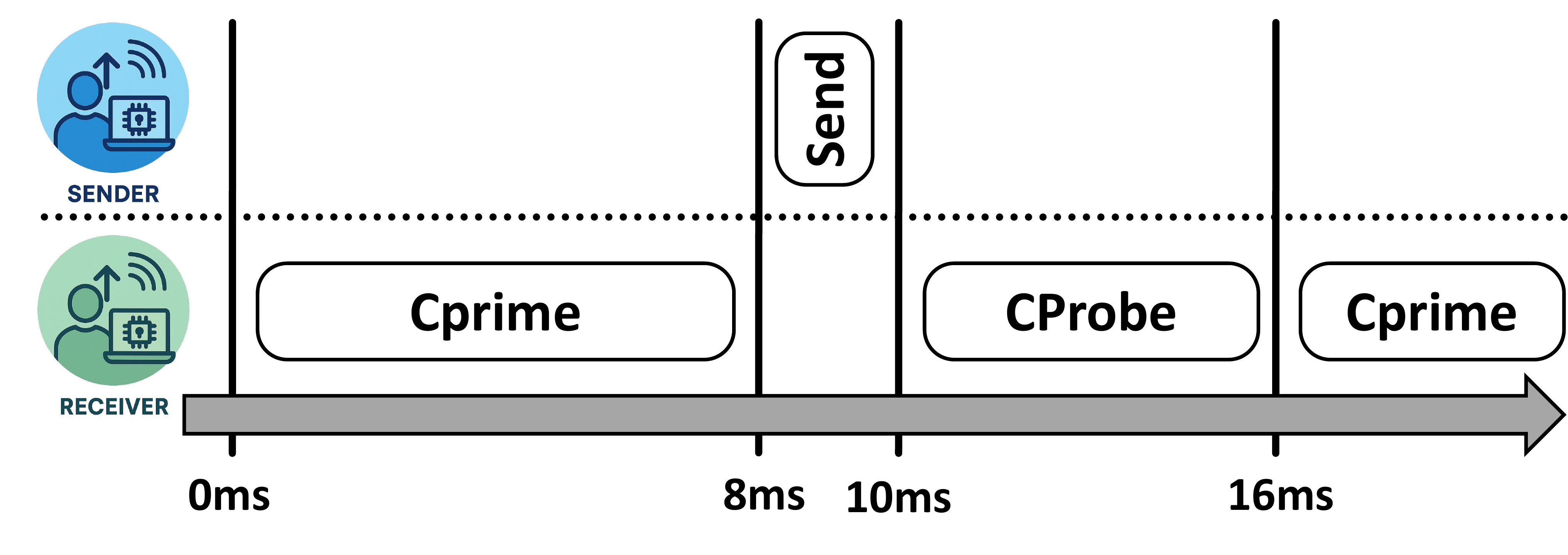}
        \caption{Cprime}
        \label{fig:covert_flow_cprime}
    \end{subfigure}
    \begin{subfigure}{\linewidth}
        \centering
        \includegraphics[width=0.8\linewidth]{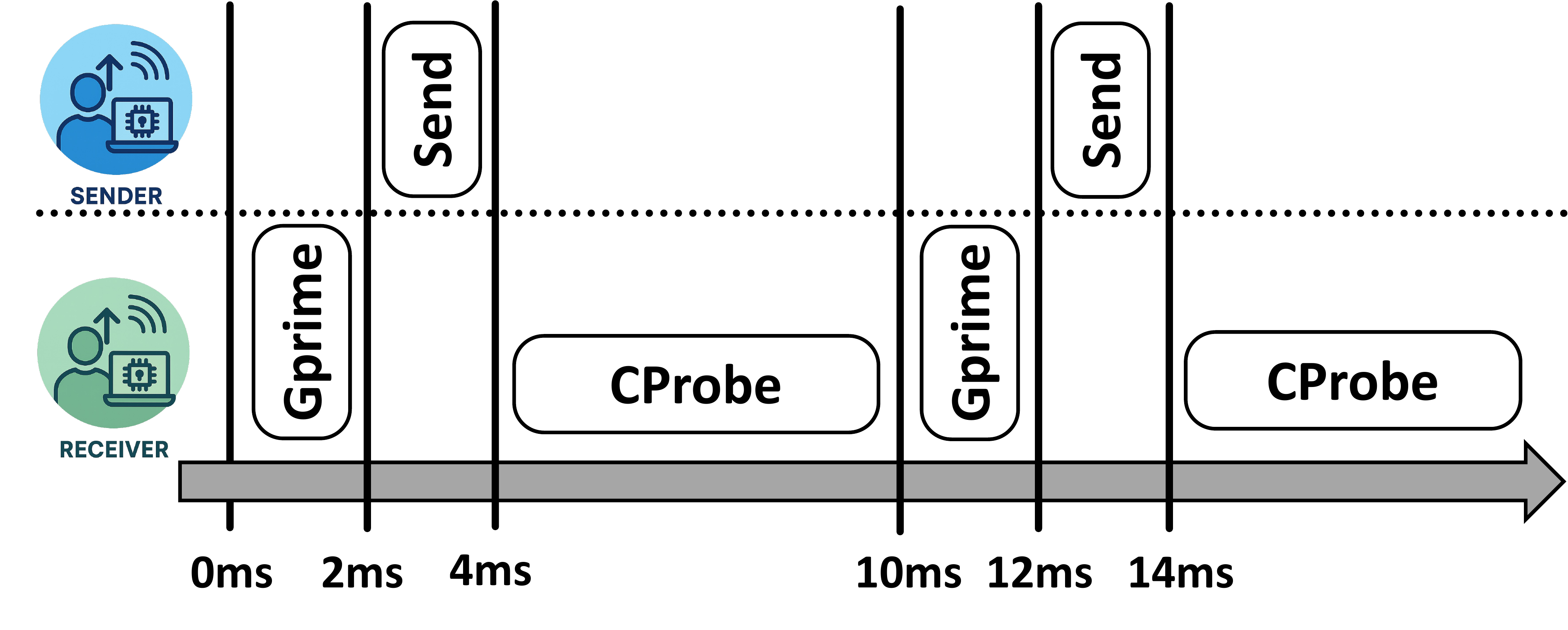}
        \caption{Gprime}
        \label{fig:covert_flow_gprime}
    \end{subfigure}
    \caption{Covert-channel communication cycle with (a)~\textsc{CPrime} and (b)~\textsc{GPrime} (shorter priming).}
    \label{fig:covert_flow}
\end{figure}
\fi

%% file: txt/5.tex
\section{End-to-end Attacks and Experimental Results}
\label{sec:results}

In this section, we demonstrate that our  CPU-to-GPU \textsc{Prime}+\textsc{Probe} side-channels enable practical privacy attacks that recover secrets of GPU workloads. In Apple Silicon's unified memory architecture, the CPU and GPU share the same physical DRAM and SLC. By monitoring the set-level SLC footprints of the GPU victim workload via our side-channels, the attacker infers the victim's runtime data accesses, thereby leaking the secret. 

Our primary attacks build on \textsc{CPrime}+\textsc{CProbe} and run entirely from an unprivileged CPU process, without GPU access. We also report alternative attacks based on \textsc{GPrime}+\textsc{CProbe} that achieve higher speed and lower noise, but require the attacker to launch GPU kernels via the Apple Metal API.

We first describe the side-channel setup common to both attacks in Section~\ref{4_5}, including noise-reduction methods for full-SLC monitoring and the three-phase workflow of \emph{profiling}, \emph{trace collection}, and \emph{secret recovery}. We then instantiate this workflow for two privacy attacks on GPU victim workloads: a \emph{GNN edge recovery attack} in Section~\ref{GNN} and an \emph{LLM input and output recovery attack} in Section~\ref{LLM}.

\subsection{Side-channel Attack Setup}\label{4_5}

The attacker observes the SLC footprint of a victim GPU kernel, which we call the \textit{target kernel}, by priming and probing the SLC around each kernel invocation. Unless stated otherwise, experiments use \textsc{CPrime}+\textsc{CProbe} as the default configuration. 

\begin{figure}[h]
    \centering
    \includegraphics[width=0.95\linewidth]{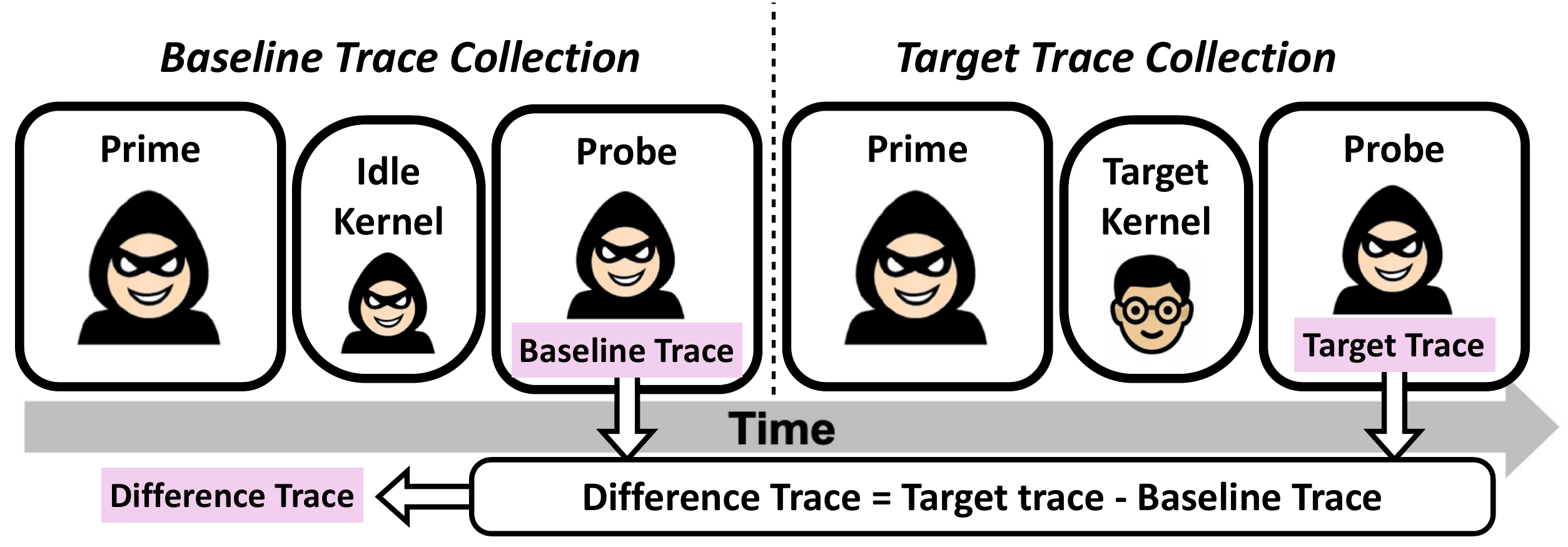}
    \caption{\textit{Differential tracing to account for system noise.}}
    \label{fig4_5}
\end{figure}

\noindent\textbf{Differential tracing.}
We observe that even in the absence of a victim workload, a small number of cache lines (typically 1 to 4 per set) are evicted by persistent background activities such as display-related work, and these evictions are stable across runs. In addition, when the victim kernel is running, it may leave SLC cache footprints unrelated to the data accesses we aim to observe. To isolate the evictions that the victim's target data accesses actually cause, we adopt \emph{differential tracing}, which collects two traces and takes their difference as the signal of interest. Figure~\ref{fig4_5} illustrates the procedure. In the \emph{Baseline Trace Collection} phase, the attacker primes the SLC, runs an \emph{idle kernel}, and probes to obtain the baseline trace. The idle kernel invokes the same library function as the victim but excludes the target data accesses (specific setup given per case study in Section~\ref{sec:results}). The \emph{Target Trace Collection} phase repeats the same prime-run-probe sequence with the actual victim kernel to obtain the target trace. Both traces are \textsc{CProbe} outputs (integer arrays of length 4{,}096, described in Section~\ref{sec:cprobe}). Subtracting the baseline from the target trace removes both persistent system noise and victim-side irrelevant activity, leaving only the data access activity of interest. We use differential tracing throughout all subsequent attacks in Section~\ref{sec:results}.

While differential tracing cancels persistent background noise, we find that additional time-varying noise remains on top of this baseline and requires further mitigation. We counter this noise with two methods that restore the decoding reliability: multi-run averaging and cache-set aggregation.

\begin{figure}[h]
\centering
\begin{minipage}[t]{0.47\linewidth}
\centering
\includegraphics[width=\linewidth]{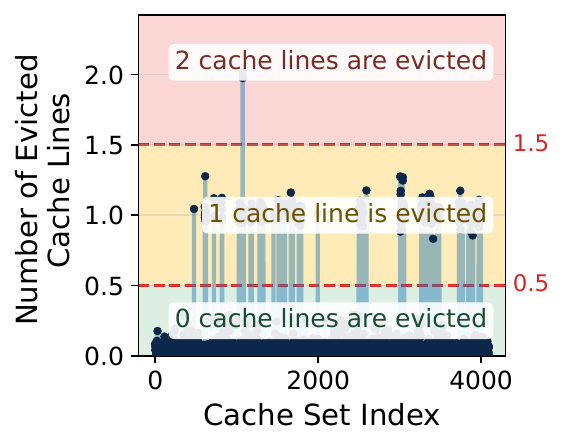}
\captionof{figure}{\textit{Averaged side-channel trace (1000 runs) for 4096 cache sets.}}
\label{fig451}
\end{minipage}\hspace{0.1cm}
\begin{minipage}[t]{0.47\linewidth}
\centering
\includegraphics[width=\linewidth]{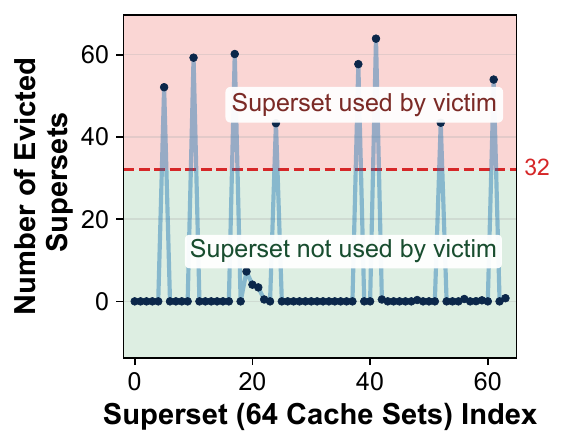}
\captionof{figure}{\textit{Aggregated single-trial side-channel trace for 64 supersets.}}
\label{fig452}
\end{minipage}
\end{figure}

\noindent\textbf{Method 1: Multi-run averaging.}
The first method reduces measurement variance by averaging the side-channel traces. Each run records a per-set SLC-miss count, an integer between 0 and 16. Averaging these counts across runs preserves the victim's footprint while smoothing out run-to-run fluctuations. Figure~\ref{fig451} shows an averaged \textsc{CPrime}+\textsc{CProbe} trace collected while the victim runs a Graph Neural Network (GNN) kernel (Section~\ref{GNN}). The averaged values observed in this trace cluster cleanly around 0, 1, and 2 occupied lines per set, and we separate these clusters using midpoint thresholds of 0.5 and 1.5. To quantify accuracy, we define \emph{undercounting} and \emph{overcounting} as the fractions of cache sets whose inferred occupancy falls below or above the ground truth. Table~\ref{tab:case1_err} (Columns 2-4) reports these rates alongside the number of averaging runs. \textsc{CPrime}+\textsc{CProbe} needs 4{,}000 runs to match the error rate that \textsc{GPrime}+\textsc{CProbe} achieves in 1{,}000 runs, because the two-stage CPU priming phase is inherently noisier than direct GPU priming.

\begin{table}[h]
\footnotesize
\centering
\caption{Accuracies with the two mitigations}
\label{tab:case1_err}
\begin{tabular}{lccc|ccc}
\toprule
Priming & \multicolumn{3}{c|}{Method1} &\multicolumn{3}{c}{Method2} \\\cline{2-7}
Method & Undercount & Overcount & \#Runs & FNR & FPR & \#Supersets\\\midrule
\textsc{CPrime} & 0.04\% & 0.53\% & 4000 & 0.17\% & 2.63\% & 64\\
\textsc{GPrime} & 0.02\% & 0.37\% & 1000 & 0.08\% & 1.03\% & 64\\
\bottomrule
\end{tabular}
\end{table}

\noindent\textbf{Method 2: Cache-set aggregation (superset).}
The second method trades per-set resolution for a cleaner single-trial signal. Instead of treating each of the 4{,}096 SLC sets independently, we partition them into equal-size groups, called \textit{superset}. For each \textit{superset} we report the aggregate number of victim-occupied lines. This aggregation is effective whenever each candidate victim data element is a multiple of (a power-of-two) cache lines. Figure~\ref{fig452} shows a single-trial \textsc{CPrime}+\textsc{CProbe} trace with 64 supersets of 64 sets each. An accessed superset reports a value close to 64, one per set, while an untouched superset stays near 0, so a midpoint threshold of 32 cleanly separates the two outcomes. Because aggregation only reveals whether a superset is accessed, we generate a binary access trace. Table~\ref{tab:case1_err} (Columns 5-7) reports the corresponding False-Negative Rate (FNR) and False-Positive Rate (FPR) over 64 supersets, with both priming strategies keeping single-trial errors below 3\%.

\if false 
\begin{table}[h]
\small
\centering
\caption{Accuracy of Method~2 (superset aggregation).}
\label{tab:case2_err}
\begin{tabular}{lccc}
\toprule
Priming Method & FNR & FPR & \#Supersets \\
\midrule
\textsc{CPrime} & 0.17\% & 2.63\% & 64 \\
\textsc{GPrime} & 0.08\% & 1.03\% & 64 \\
\bottomrule
\end{tabular}
\end{table}
\fi 

\begin{figure*}[t]
    \centering
    \includegraphics[width=\linewidth]{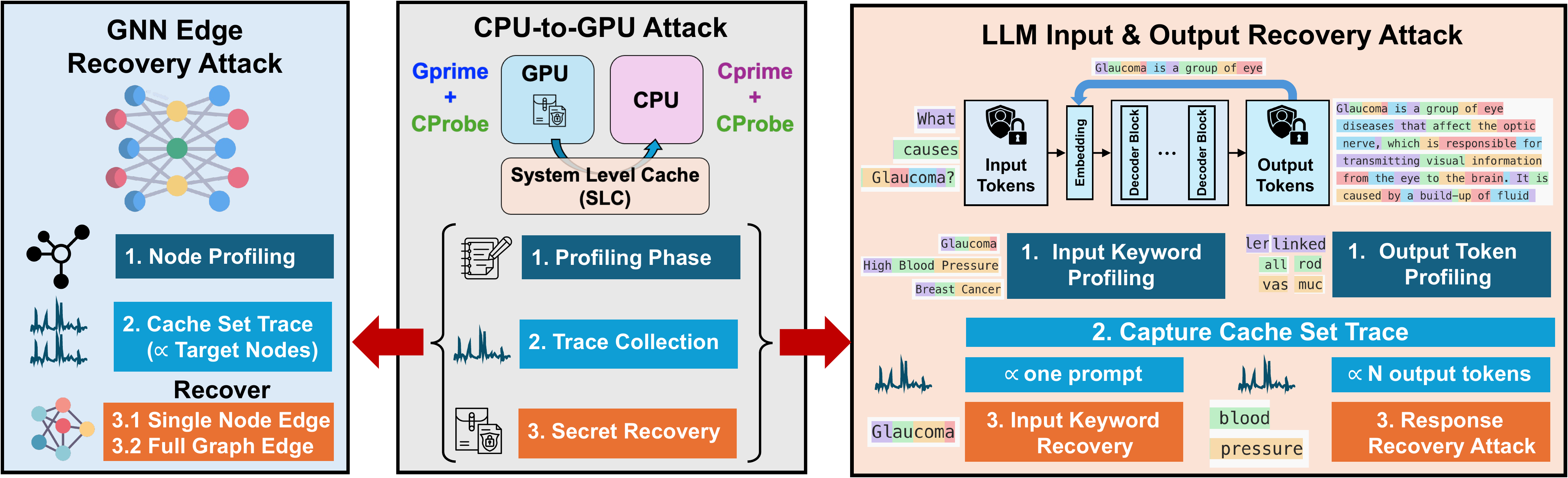}
    \caption{Proposed CPU-to-GPU cache side-channel attack workflow using \textsc{CPrime}+\textsc{CProbe} and \textsc{GPrime}+\textsc{CProbe}, with its application to GNN edge recovery and LLM input/output recovery attacks.}
    \label{fig:Attack-setup}
\end{figure*}

\noindent\textbf{Three-phase attack methodology.}
Both the GNN and LLM attacks share a common three-phase workflow with their respective noise-reduction method. We model the victim's GPU workload as accessing items drawn from a finite \emph{Target Item List} $\mathcal{L}=\{i_1,i_2,\dots,i_N\}$, where each item corresponds to a data structure in memory such as a feature vector, a lookup-table entry, or a matrix row. Figure~\ref{fig:Attack-setup} illustrates the general three-phase workflow along with its two instantiations: the GNN edge recovery attack (left panel) and the LLM input and output recovery attack (right panel). The three phases are as follows.

\begin{enumerate}[noitemsep,topsep=4pt,leftmargin=13pt]
    \item \textbf{Profiling.} The attacker profiles each item $i_i$ individually to build an \emph{Access Mapping Matrix} $\mathcal{M}^{N \times S}$ for the entire list  $\mathcal{L}$, where $S$ is the number of  SLC cache sets (Method~1) or supersets (Method~2). Under Method~1, $\mathcal{M}_{i,j} \in \{0, 1, \ldots, 16\}$ is the cache line occupancy of item $i_i$ in set $j$. Under Method~2, $\mathcal{M}_{i,j} \in \{0, 1\}$ indicates superset $j$ no-access versus access by item $i_i$. 
    \item \textbf{Trace collection.} The attacker obtains an SLC side-channel trace for each victim kernel invocation, denoted as an \emph{Observed Access Vector} $\mathbf{v} \in \mathbb{Z}^{S}$. Under Method~1, $v_j \in \{0, 1, \ldots, 16\}$ is the number of victim-occupied lines in cache set $j$. Under Method~2, $v_j \in \{0, 1\}$ is the observed superset $j$ access flag.
    \item \textbf{Secret recovery.} The attacker compares $\mathbf{v}$ against the rows of $\mathcal{M}$ to infer which items in $\mathcal{L}$ the victim accessed during that invocation. The specific comparison algorithm depends on the attack and is detailed in Sections~\ref{GNN} and~\ref{LLM}.
\end{enumerate}

\subsection{GNN Edge Recovery Attack}\label{GNN}

Graph Neural Networks (GNNs) are commonly used in recommendation systems and social networks. For node inference, a target node's representation is computed by aggregating the features of its neighbor nodes. In applications such as social networks and biomedical analysis, the graph's edge structure encodes proprietary or privacy-critical relationships. Because the neighbors accessed during inference directly reflect these edges, the memory access pattern becomes a high-value target for privacy attacks.

\noindent\textbf{Threat model.} A GNN is deployed on an Apple M-series SoC using PyTorch with Metal GPU acceleration. The attacker is a co-resident, unprivileged CPU process that can issue arbitrary node queries to the model for inference, a common chosen-query setting~\cite{zhuang2024unveiling, podhajski2026stealing}. The attacker's goal is to recover the full set of graph edges from side-channel observations collected during these queries. We evaluate both \textsc{CPrime}+\textsc{CProbe} and \textsc{GPrime}+\textsc{CProbe} attacks.

\noindent\textbf{Deployment characteristics.} When the model is deployed, PyTorch allocates the feature vectors of all graph nodes in the unified memory via Apple's Metal API. During inference on a target node, the model invokes \texttt{Torch.index\_select()} to gather its neighbors' feature vectors into a temporary tensor for aggregation. This gather operation is the source of side-channel leakage: each neighbor's feature vector occupies a distinct region of memory, and the GPU's read operations leave clear footprints on the SLC that a CPU-side attacker can monitor. Accordingly, the idle kernel used for differential tracing (Section~\ref{4_5}) invokes the \texttt{Torch.index\_select()} call with an empty index set.

\noindent\textbf{Attack workflow.} We instantiate the general three-phase attack workflow for GNN edge recovery, treating each graph node's feature vector as an item $i_i$ in the Target Item List $\mathcal{L}$. The left panel of Figure~\ref{fig:Attack-setup} illustrates the workflow.

\begin{enumerate}[noitemsep,topsep=4pt,leftmargin=13pt]
    
    \item \textbf{Profiling.} The attacker constructs a binary Access Mapping Matrix $\mathcal{M}_G \in \{0,1\}^{N \times 4096}$, where each row corresponds to one graph node feature vector,  recording which  SLC cache sets the feature vector maps to. Because macOS does not expose the physical addresses of GPU-resident tensors to user-space processes, the attacker cannot directly apply the SLC indexing function from Section~\ref{set_index} and instead profiles $\mathcal{M}_G$ empirically with the full SLC priming and probing. 

    \item \textbf{Trace collection.} During a target-node query, the GNN kernel accesses all of that node's neighbors' feature vectors in a single batch, so the resulting Observed Access Vector $\mathbf{v} \in \mathbb{Z}^{4096}$ is the superposition of all neighbors' SLC footprints. Each entry $v_j \in \{0, 1, \ldots, 16\}$ counts the cache lines  in set $j$ the victim accesses. We collect $\mathbf{v}$ with Method~1 (multi-run averaging) to reduce the system noise.

    \item \textbf{Secret recovery.} The attacker compares $\mathbf{v}$ against the rows of $\mathcal{M}_G$ to infer which nodes the victim accessed during the target-node query, and each accessed node yields one edge of the target node. We measure the recovery quality using \textit{precision} (fraction of reported edges that are true) and \textit{recall} (fraction of true edges that are recovered). The recovery proceeds in two stages as described next.
\end{enumerate} 

\noindent\textbf{Single-node edge recovery.} Given the access vector $\mathbf{v}$ for a target node $n$, the attacker recovers its neighbors by solving a \emph{set cover} problem over the rows of $\mathcal{M}_G$: identifying the minimal collection of node profiles whose combined cache-set footprint matches $\mathbf{v}$. The attacker first filters $\mathcal{M}_G$ to a candidate set $\hat{\mathcal{C}}$ of profiles that are element-wise dominated by $\mathbf{v}$, namely those satisfying $\mathcal{M}_G[i,j] \leq \mathbf{v}[j]$ at every cache-set index $j$. The attacker then prunes $\hat{\mathcal{C}}$ down to the intersection of all size-$\hat{m}$ subsets whose combined footprint matches $\mathbf{v}$, where $\hat{m}$ is the neighbor count estimated from the total occupancy in $\mathbf{v}$. Pruning is invoked only when $\hat{m}$ falls below a dataset-specific trustworthiness threshold $T$ (Table~\ref{tab1}), and the attacker otherwise returns $\hat{\mathcal{C}}$ unchanged. The full algorithm, including the derivation of $T$ and the additional fallback paths, is given in Appendix~\ref{app:single-node-recovery}.

\noindent{\textbf{Full-graph edge recovery.}} After recovering a candidate neighbor set for each node, the attacker reconstructs the entire edge set by aggregating these per-node candidates. Because the graph is undirected, a true edge $(u, v)$ must appear in the candidate sets of both $u$ and $v$, so the attacker keeps only edges confirmed from both endpoints. False positives rarely confirm bidirectionally, so this check improves precision while preserving high recall.

\noindent\textbf{Evaluation.} We evaluate the GNN edge recovery attack on a Graph Convolutional Network (GCN)~\cite{van2017graph} over five benchmark datasets: \texttt{Cora}, \texttt{Citeseer}, \texttt{Pubmed}, \texttt{Computers}, and \texttt{Photo}~\cite{yang2016revisiting, shchur2018pitfalls}. Table~\ref{tab1} summarizes the dataset characteristics and reports the attack accuracy under both \textsc{CPrime}+\textsc{CProbe} and \textsc{GPrime}+\textsc{CProbe} in terms of precision and recall.

\definecolor{cprimecol}{RGB}{217,234,250}
\definecolor{gprimecol}{RGB}{253,233,217}

\begin{table*}[t]
\centering
\caption{Datasets and GNN edge recovery attack results, reported as precision/recall (P/R). P is the percentage of reported edges that are true, and R is the percentage of true edges recovered. Columns are shaded by attack: \colorbox{cprimecol}{\textsc{CPrime}} and \colorbox{gprimecol}{\textsc{GPrime}}.}
\label{tab1}
\small
\setlength{\tabcolsep}{3.5pt}
\begin{tabular}{cccccc
  >{\columncolor{cprimecol}}c >{\columncolor{gprimecol}}c
  >{\columncolor{cprimecol}}c >{\columncolor{gprimecol}}c
  >{\columncolor{cprimecol}}c >{\columncolor{gprimecol}}c}
\toprule
\multirow{2}{*}{Dataset}
 & \multirow{2}{*}{\#Nodes (N)}
 & \multirow{2}{*}{\#Edges}
 & \multirow{2}{*}{\#Feat.}
 & \multirow{2}{*}{\makecell[c]{Avg. \# of cache\\lines per node ($\bar{f}$)}}
 & & \multicolumn{2}{c}{Threshold $T$}
   & \multicolumn{2}{c}{Single-Node Recovery}
   & \multicolumn{2}{c}{Full-Graph Recovery} \\
\cmidrule(lr){7-8} \cmidrule(lr){9-10} \cmidrule(lr){11-12}
 & & & & &
 & \textsc{CPrime} & \textsc{GPrime}
 & \textsc{CPrime}\,(P/R) & \textsc{GPrime}\,(P/R)
 & \textsc{CPrime}\,(P/R) & \textsc{GPrime}\,(P/R) \\
\midrule
Cora     & 2{,}708  & 10{,}556  & 1{,}433 & 46.78   & & 60 & 95  & 84.3 / 97.2 & 90.6 / 98.8 & 98.2 / 94.2 & 99.9 / 97.3 \\
Citeseer & 3{,}327  & 9{,}104   & 3{,}703 & 117.72  & & 74 & 105 & 91.6 / 94.2 & 95.2 / 97.3 & 97.8 / 90.5 & 99.2 / 94.1 \\
Pubmed   & 19{,}717 & 88{,}648  & 500     & 17.63   & & 28 & 40  & 82.9 / 99.1 & 83.2 / 99.6 & 99.8 / 97.9 & 99.9 / 98.8 \\
Computers & 13{,}752 & 491{,}722 & 767     & 25.97   & & 53 & 93  & 30.8 / 98.5 & 33.2 / 99.2 & 86.6 / 96.8 & 87.2 / 98.6 \\
Photo    & 7{,}650  & 238{,}162 & 745     & 25.29   & & 36 & 58  & 46.6 / 98.5 & 48.3 / 99.3 & 93.8 / 97.2 & 94.9 / 98.6 \\
\bottomrule
\end{tabular}
\end{table*}

With \textsc{CPrime}+\textsc{CProbe}, the CPU-only attacker achieves strong results. Single-node recall is consistently high, typically above 94\% and often exceeding 98\%, which reflects the low undercounting rate of the side-channel. Single-node precision, however, varies with graph density. Sparse graphs (\texttt{Cora}, \texttt{Citeseer}, \texttt{Pubmed}) keep precision in the 83-92\% range because nodes have few neighbors and per-neighbor cache footprints rarely overlap. Dense graphs (\texttt{Computers}, \texttt{Photo}) drop in precision because the larger neighborhoods produce extensive cache-set overlap and inflate $\mathcal{C}$ with spurious candidates. The full-graph recovery stage substantially closes this gap, raising \textsc{CPrime}+\textsc{CProbe} precision above 86\% on all datasets while keeping recall above 90\%. \textsc{GPrime}+\textsc{CProbe} attacks achieve higher precision and recall because GPU-based priming yields a cleaner SLC state. Full-graph precision then exceeds 99\% on \texttt{Cora}, \texttt{Citeseer}, and \texttt{Pubmed}.

\subsection{LLM Privacy Attacks}\label{LLM}

Large Language Models (LLMs), particularly decoder-only architectures such as GPT and Llama, are now the dominant paradigm for language-based inference. A decoder-only model consists of an embedding layer, a stack of transformer decoder blocks, and a projection layer. The embedding layer looks up a table in memory for the given input (prompt token or the previously output response token),  whose rows store the fixed-dimensional embedding of each token in the vocabulary. Because these reads are input-dependent and span both the \textit{prefill} and \textit{decoding} phases, the embedding layer forms a natural side-channel attack surface revealing private queries, proprietary data, or confidential inferences. The embedding layer aligns with the \emph{Target Item List} abstraction in Section~\ref{4_5}: each token embedding corresponds to an item $i_i$, and the attacker's goal is to infer which embeddings the victim accesses by observing SLC eviction patterns.

\noindent\textbf{Threat model.} The victim deploys a decoder-only LLM on the GPU of an Apple M-series SoC using PyTorch with Metal GPU acceleration. The attacker is a co-resident, unprivileged background process on the same system. Although the LLM model is provided as a shared library on the platform, the victim runtime information, including the victim's input prompts and the generated responses, is kept private with process isolation. The attacker only relies on the SLC side-channel to observe the embedding-layer footprints left by the victim's execution. 

\noindent\textbf{Deployment characteristics.} PyTorch loads the token embedding table into unified memory via Apple's Metal API, and the embedding layer is invoked on every forward pass. The GPU's read operations leave clear footprints on the SLC that a CPU-side attacker can monitor, making the embedding-layer lookup the \emph{target kernel} of our attack. Accordingly, the \emph{idle kernel} used for differential tracing (Section~\ref{4_5}) invokes the same embedding layer with an empty input token set, so it exercises the same code path as the target kernel without performing any token-driven embedding lookups.

We evaluate two applications, in which the victim queries the LLM about one of the 1{,}000 most common diseases (drawn from the \textsc{MedQuad}~\cite{medquad2023} prompt dataset) or one of the top 100 publicly traded stock tickers (drawn from the \textsc{FinancialQA}~\cite{saeedian2023financialqa} prompt dataset). In each case, the attacker maintains a matching domain-specific dictionary of disease names or stock tickers.

\noindent\textbf{Attack overview.} We instantiate the three-phase attack workflow of Section~\ref{4_5} in two complementary attacks. The first, \emph{input keyword recovery}, targets the \textit{prefill} phase, in which the model processes the prompt and looks up embeddings for all input tokens in a single batch, aiming to identify sensitive keywords or topics in the user's prompt. The second, \emph{response recovery}, targets the \textit{decoding} phase, in which the model generates the output autoregressively and accesses the embedding layer once per step to look up the previously generated token, aiming to reconstruct the model's output token-by-token. The right panel of Figure~\ref{fig:Attack-setup} shows both attacks.

\noindent\textbf{Superset setting.} Unlike the GNN attack, which monitors all 4{,}096 individual SLC cache sets, the LLM attack groups cache sets into \emph{supersets} and monitors at a coarser granularity, following Method~2 of Section~\ref{4_5} to reduce side-channel noise. This aggregation is applicable to LLM inference because embedding tables have a highly regular memory layout. Each token embedding is a fixed-size vector whose dimension is typically a power of two (e.g., 1024 or 2048), and embedding layers are commonly page-aligned so that the first vector starts at page offset zero~\cite{zhang2024tinyllama, radford2019language}. Each embedding, therefore, occupies a contiguous, aligned region of physical memory.

The alignment of embeddings interacts predictably with the SLC set-indexing functions. Figure~\ref{fig:hashheatmap} shows that on Apple M1, for the lower 6 SLC set index bits, each is directly obtained from one address bit (within 7--12). Now consider an embedding of size $2^m$ bytes with $7 < m < 12$.  For its memory block addresses, the address bits below $m$ vary across cache-line blocks, whereas bits $m$ and above are fixed. Applying the indexing functions, only the lower $m-7$ SLC index bits sweep through all values for the embedding, while the higher index bits stay constant. Each embedding thus maps to a contiguous, power-of-two-aligned block of $2^{m-7}$ cache sets that does not overlap with any other embedding's block. These blocks are supersets that partition the 4{,}096 SLC sets into a lower dimension, with the side-channel length reduced from 4{,}096 to 128 or 256, depending on the embedding size.

We profile two LLMs, TinyLlama~\cite{zhang2024tinyllama} and GPT-2 Medium~\cite{radford2019language}. Table~\ref{tab3} summarizes their structural characteristics, the token embedding memory size, cache-line count per token, and the resulting number of supersets.

\begin{table}[h]
\centering
\caption{Memory and SLC Usage of LLMs Token Embedding.}
\label{tab3}
\resizebox{\columnwidth}{!}{%
\begin{tabular}{lcc|ccc}
\toprule
\textbf{Model} & \textbf{\#Param} & \textbf{\makecell{Vocab \\ Size}} & \textbf{\makecell{Memory \\ /Token}} & \textbf{\makecell{\#Lines \\ /Token}} & \textbf{\makecell{\# \\ Supersets}} \\
\midrule
TinyLlama\cite{zhang2024tinyllama}      & 1.1B   & 32,000  & 4KB & 32 & 128 \\
GPT-2 Medium\cite{radford2019language}  & 0.35B  & 50,257  & 2KB & 16 & 256 \\
\bottomrule
\end{tabular}}
\end{table}

\subsubsection{Input Keyword Recovery Attack.} This attack targets the \textit{prefill} phase of LLM inference to identify the sensitive keyword in the user's prompt (e.g., a disease name or stock ticker) by monitoring the SLC evictions caused by embedding-layer lookups.
\begin{enumerate}[noitemsep,topsep=4pt,leftmargin=13pt]
    \item \textbf{Profiling.} The attacker profiles the SLC superset access pattern for each keyword in the domain-specific dictionary, treating each keyword as a sequence of one or more tokens. Profiling all $n$ keywords produces a binary matrix $\mathcal{M}_K \in \{0,1\}^{n \times S}$, where $S$ is the number of supersets and each row records which supersets a given keyword maps to. When the tokens of a keyword map to disjoint supersets, the row's active-superset count equals the token count. Most keywords in our dictionaries span 3-8 tokens.
    \item \textbf{Trace collection.} The attacker monitors the SLC supersets during the victim's prefill phase and records a binary vector $\mathbf{v}_\text{input} \in \{0,1\}^{S}$ in which each entry indicates whether the corresponding superset was accessed. Because the prompt contains additional tokens beyond the target keyword (function words, punctuation, and so on), $\mathbf{v}_\text{input}$ is generally denser than the keyword's own superset profile.
    \item \textbf{Secret recovery.} The attacker compares $\mathbf{v}_\text{input}$ against each row of $\mathcal{M}_K$ and flags every keyword whose superset profile is fully contained in $\mathbf{v}_\text{input}$ as a candidate. Ideally, this yields a unique candidate, since each prompt contains exactly one keyword from the dictionary. However, supersets activated by non-keyword tokens may also satisfy containment for unrelated keywords, introducing false positives. When multiple candidates remain, the attacker selects one randomly.
\end{enumerate}

\begin{table}[ht]
\centering
\caption{Accuracy of LLM Input Keyword Recovery}
\label{tab:input_recovery_accuracy}
\setlength{\tabcolsep}{6pt}
\renewcommand{\arraystretch}{1}
\begin{tabular}{lcccc}
\toprule
& \multicolumn{2}{c}{TinyLlama} & \multicolumn{2}{c}{GPT-2 Medium} \\
\cmidrule(lr){2-3}\cmidrule(lr){4-5}
Dataset & \textsc{CPrime} & \textsc{GPrime} & \textsc{CPrime} & \textsc{GPrime} \\
\midrule
MedQuad      & 76.3\% & 91.5\% & 78.2\% & 93.4\% \\
FinancialQA  & 80.5\% & 94.2\% & 81.2\% & 94.8\% \\
\bottomrule
\end{tabular}
\end{table}

Table~\ref{tab:input_recovery_accuracy} reports keyword recovery accuracy, 76--81\% across the two datasets and two models for \textsc{CPrime}+\textsc{CProbe},  rising to 91--95\% for \textsc{GPrime}+\textsc{CProbe}, reflecting the lower noise floor of \textsc{GPrime}.

\subsubsection{Response Recovery Attack} Unlike the input keyword recovery attack, which targets the \textit{prefill} phase, response recovery operates token-by-token during the autoregressive \textit{decoding}, which yields cleaner per-token side-channel observations and requires a simpler recovery algorithm. The attack uses the same three-phase workflow.

\begin{enumerate}[noitemsep,topsep=4pt,leftmargin=13pt]
    \item \textbf{Profiling.}  The attacker profiles individual token embeddings rather than multi-token keywords. As the full vocabularies are prohibitively large (see Table~\ref{tab3}), we restrict profiling to the 3{,}000 most frequent tokens, which cover over 95\% of tokens in our evaluation prompts. Profiling produces a binary matrix $\mathcal{M}_O \in \{0,1\}^{T \times S}$, where each row records the superset mapping of one of the $T$ profiled tokens across $S$ supersets.

    \item \textbf{Trace collection.} At each decoding step, the embedding layer looks up exactly one token embedding and activates a single SLC superset, which the attacker reads as the superset with the maximum observed access value. For a response of $r$ output tokens, the attacker therefore collects $r-1$ traces. This direct superset identification achieves high accuracy on both models under both priming strategies: 99.2\% (\textsc{CPrime}) and 99.7\% (\textsc{GPrime}) on TinyLlama, and 98.8\% (\textsc{CPrime}) and 99.5\% (\textsc{GPrime}) on GPT-2 Medium. GPT-2 Medium scores slightly lower than TinyLlama because its larger embedding table spans more supersets, raising the chance of superset-level ambiguity.

    \item \textbf{Secret recovery.} Although the attacker identifies the accessed superset at each decoding step with high accuracy, $\mathcal{M}_O$ has only 128 or 256 supersets but covers about 3{,}000 profiled tokens, so each superset maps to many tokens. Each observed superset, therefore, yields a candidate set $\mathcal{C}_t$ rather than a unique token. To resolve this ambiguity, the attacker uses a local copy of the same LLM as a language model. Given the prefix of already-recovered tokens, the attacker runs the LLM to obtain logits over the full vocabulary, restricts the softmax to the candidates in $\mathcal{C}_t$, and selects the most probable token.
\end{enumerate}

\emph{Bootstrapping the initial output tokens.} 
Initially, the prefix is either empty or too short to disambiguate early candidates. To handle this, the attacker jointly recovers the first three tokens by enumerating all tuples $(x_1, x_2, x_3)$ with $x_i \in \mathcal{C}_i$ and scoring each tuple as:
\begin{equation}
\mathrm{Score}(x_1,x_2,x_3) = P_{\text{LLM}}(x_1 \mid \emptyset) \cdot P_{\text{LLM}}(x_2 \mid x_1) \cdot P_{\text{LLM}}(x_3 \mid x_1, x_2),
\end{equation}
where each probability term comes from running the LLM on the corresponding prefix. The tuple with the highest score becomes the starting prefix.

The choice of three jointly recovered tokens trades accuracy against search cost. Larger windows provide the language model with more context, but they also exponentially increase the search space with respect to the number of positions. In our experiments, going beyond three tokens yields only marginal accuracy improvements while increasing the search cost by a factor of tens, so three is the smallest window that captures the accuracy benefit.

\emph{Iterative decoding for the remaining tokens.} Once the first three tokens are fixed, the attacker recovers each subsequent token $t \geq 4$ greedily by selecting the candidate in $\mathcal{C}_t$ with the highest probability under the recovered prefix:
\begin{equation}
\hat{x}_t = \arg\max_{x \in \mathcal{C}_t}\; P_{\text{LLM}}(x \mid \hat{x}_1, \ldots, \hat{x}_{t-1}).
\end{equation}
\begin{table}[ht]
\centering
\caption{Accuracy of LLM Output Response Recovery}
\label{tab:output_recovery_accuracy}
\setlength{\tabcolsep}{6pt}
\renewcommand{\arraystretch}{1.05}
\begin{tabular}{lcccc}
\toprule
& \multicolumn{2}{c}{TinyLlama} & \multicolumn{2}{c}{GPT-2 Medium} \\
\cmidrule(lr){2-3}\cmidrule(lr){4-5}
Dataset & \textsc{CPrime} & \textsc{GPrime} & \textsc{CPrime} & \textsc{GPrime} \\
\midrule
MedQuad      & 70.5\% & 75.7\% & 79.2\% & 83.6\% \\
FinancialQA  & 73.4\% & 78.7\% & 85.3\% & 88.9\% \\
\bottomrule
\end{tabular}
\end{table}

Table~\ref{tab:output_recovery_accuracy} reports the response recovery accuracy, defined as the fraction of output tokens correctly recovered across the response. \textsc{CPrime}+\textsc{CProbe} attacks recover 70--85\% of output tokens correctly across the two models and two datasets. \textsc{GPrime}+\textsc{CProbe} attacks improve the
accuracy by 4--5\%. Because each decoded token conditions all subsequent predictions, early errors propagate through the autoregressive chain, leading to cascading mispredictions. This error amplification explains why the \textsc{CPrime}-versus-\textsc{GPrime} gap is larger here than at the superset level: although \textsc{CPrime}'s superset accuracy is less than 1\% below \textsc{GPrime}'s, the small per-step gap compounds across sequential decoding. GPT-2 Medium scores higher than TinyLlama on response recovery despite a slightly lower superset accuracy, because its larger embedding table distributes tokens across more supersets, yielding fewer candidates per superset and making disambiguation more reliable.

%% file: txt/6.tex
\section{Conclusion and Future Work}
\label{sec:conclusion}

This paper presents SLAC, the first fine-grained Prime+Probe CPU-to-GPU cache side-channel attack on Apple Silicon. By reverse-engineering the Apple M1 SLC, including its 12-bit set-indexing hash functions, we constructed two side-channel techniques.  \textsc{CPrime} +\textsc{CProbe} lets an unprivileged CPU process observe GPU victim activity, and \textsc{GPrime}+\textsc{CProbe} uses GPU-side priming for a higher attack speed. The corresponding covert channels achieved throughputs of 62.5~Kbps and 400~Kbps, respectively. We demonstrated two end-to-end privacy attacks based on the new side-channels: a GNN edge recovery attack with over 90\% accuracy across five benchmark datasets, and an LLM privacy attack on TinyLlama and GPT-2 Medium that recovered input keywords with up to 94.8\% accuracy and output tokens with up to 88.9\% accuracy.

The broader takeaway is that the shared SLC in Apple's unified memory architecture is a potent cross-domain attack surface, enabling fine-grained leakage of sensitive GPU workloads to an unprivileged CPU process. As heterogeneous CPU-GPU SoCs increasingly host privacy-critical ML workloads, these findings motivate careful redesign of cache hierarchies and coherence protocols to mitigate access-driven side-channels across compute domains.

\noindent\textbf{Future work.} The first direction is broadening this attack family across the Apple M-series. M1 Pro/Max and M2 share the same overall architecture as M1, including separate CPU and GPU clusters, a shared SLC, and the asymmetric SLC residency that our side-channel techniques exploit, so we expect both \textsc{CPrime}+\textsc{CProbe} and \textsc{GPrime}+\textsc{CProbe} to apply with only platform-specific calibration, such as re-recovering the SLC set-indexing functions. The Apple M3 SoC, however, presents additional challenges: it introduces Dynamic Caching for GPU memory, which could affect interactions between local caches and the SLC.

The second direction is discovering cross-domain side-channels on other heterogeneous platforms. Intel iGPUs~\cite{dutta2021leaky} and AMD APUs \cite{wang2025zenleak} use non-inclusive GPU cache policies, so the residency asymmetry we exploit on Apple Silicon manifests differently on those designs and opens a complementary set of shared channels. Promising candidates include memory-controller occupancy, interconnect contention, and Prime+Probe variants tailored to each vendor's inclusiveness rules, all of which would extend cross-domain side-channel analysis to the broader heterogeneous SoC landscape.

%% file: txt/appendix.tex
\appendix
\section*{Appendix}
\setcounter{subsection}{0}
\renewcommand\thesubsection{\Alph{subsection}}

\section{Open Science}
In line with the CCS open-science policy, we release an anonymous artifact package at \url{https://anonymous.4open.science/r/SLAC_CCS2026-7FB3}. The package contains: (1) The implementation of our two side-channel techniques described in Section~\ref{sec:slc_attack}, namely \textsc{CPrime}+\textsc{CProbe} and \textsc{GPrime}+\textsc{CProbe}, (2) The trace-processing scripts from Section~\ref{4_5} that reproduce Figures~\ref{fig451} and \ref{fig452}, and (3) The attack algorithms for the GNN edge recovery attack (Section~\ref{GNN}) and the LLM input/output recovery attacks (Section~\ref{LLM}). Following the instructions in the package, evaluators can reproduce all major experiments reported in the paper.

\section{Ethical Considerations}
We have considered the ethical implications of releasing this work and have taken the following steps to balance public benefit against the risk of misuse. The vulnerability described in this paper was identified through experiments conducted entirely on hardware owned by the authors, using victim workloads that we ran ourselves, so no third-party users, third-party services, or production systems were profiled at any point. The GNN evaluations rely on standard public benchmarks, and the LLM evaluations use the public \textsc{MedQuad} and \textsc{FinancialQA} prompt datasets, which contain no personally identifiable information. Therefore, the study involved no human subjects. We believe that publishing the reverse-engineering methodology, the two side-channel techniques, and the end-to-end attacks is necessary for the community to study and ultimately mitigate this class of leakage in heterogeneous SoCs, and we therefore plan to release our findings and analysis tools publicly so that other researchers can independently verify the vulnerability and build defenses on top of it. 

\section{SLC copy verification protocol}
\label{app:slc-copy}

In this section, we investigate whether the SLC creates an extra copy when a CPU-owned block becomes shared with the GPU, using CPU-side latency measurement. We compare two sequences of accessing two buffers (each of the size of SLC) that differ only in whether the CPU first accesses a buffer (therefore warms up its local caches):

\paragraph{Sequence A: CPU warmed target}
\begin{algorithm}[h]
\caption{CPU warms target, then GPU shares it}
\label{alg:SLCExpt}
\begin{algorithmic}[1]
\State Allocate two disjoint buffers: \texttt{buffer1} (target) and \texttt{buffer2} (probe), each about the SLC size.
\State CPU sequentially accesses \texttt{buffer1} to keep it in CPU private caches.
\State GPU accesses \texttt{buffer2} to populate the SLC with \texttt{buffer2}.
\State GPU accesses \texttt{buffer1} so that the block becomes shared.
\State CPU accesses \texttt{buffer2} and records the latency.
\end{algorithmic}
\end{algorithm}

\noindent
If sharing \texttt{buffer1} created an extra SLC copy, some lines of \texttt{buffer2} would be evicted, and the CPU probe would look like a memory access (Algorithm~\ref{alg:SLCExpt}). We instead observe SLC hit latencies. This means no new SLC copy is created when the target is already present in CPU private caches; the GPU obtains the data via the coherence path, and the CPU lines remain resident.

\paragraph{Sequence B: GPU first touch of target}
\begin{algorithm}[h]
\caption{GPU brings the target into the SLC, then CPU probes}
\label{alg:SLCExptNoWarm}
\begin{algorithmic}[1]
\State Allocate \texttt{buffer1} and \texttt{buffer2} as above.
\State GPU accesses \texttt{buffer2} to populate the SLC with \texttt{buffer2}.
\State GPU accesses \texttt{buffer1} without prior CPU warmup.
\State CPU accesses \texttt{buffer2} and records the latency.
\end{algorithmic}
\end{algorithm}

\noindent
Here, the first touch to \texttt{buffer1} must insert it into the SLC (Algorithm~\ref{alg:SLCExptNoWarm}). The CPU probe to \texttt{buffer2} shows memory level latency, which is consistent with eviction of \texttt{buffer2} lines to make room for \texttt{buffer1}.

In the first sequence, the CPU loads \texttt{buffer1} into private caches, and the GPU loads \texttt{buffer2} to SLC. The GPU then accesses \texttt{buffer1}, and the CPU finally probes \texttt{buffer2} with time measurement. The latency remains at SLC hit latency, indicating that when \texttt{buffer1} becomes shared from CPU-private, no second copy is created in SLC and the GPU obtains the data directly through the coherence protocol, an unusual \textit{non-inclusive behavior} on the GPU side. In the second sequence, we omit the CPU warmup of local caches with \texttt{buffer1}, so the GPU loads \texttt{buffer2} into SLC first, followed by \texttt{buffer1}. When CPU probes \texttt{buffer2}, all accesses hit the memory, indicating that \texttt{buffer1} indeed enters the SLC and has evicted prior contents.

\section{Single-Node Edge Recovery}
\label{app:single-node-recovery}

This appendix details the single-node edge recovery procedure referenced in Section~\ref{GNN}. Throughout, we use the following notation:
\begin{itemize}[noitemsep,leftmargin=15pt,topsep=3pt]
    \item $m$: the true number of neighbors of the target node $n$;
    \item $\mathcal{C}$: the ground-truth neighbor set, with $|\mathcal{C}| = m$;
    \item $\hat{\mathcal{C}}$: the intermediate candidate set produced by the filtering step;
    \item $\hat{m}$: the neighbor count estimated from the access vector $\mathbf{v}$;
    \item $\tilde{\mathcal{C}}$: the final returned set after pruning.
\end{itemize}
The recovery proceeds in three steps: (1)~filtering all nodes to obtain $\hat{\mathcal{C}}$, (2)~estimating $\hat{m}$ and assessing its reliability via a dataset-specific trustworthiness threshold $T$, and (3)~pruning $\hat{\mathcal{C}}$ to produce $\tilde{\mathcal{C}}$.

\subsection*{Step 1: Filtering}

A node $i$ is admitted to $\hat{\mathcal{C}}$ when its profiled cache pattern $\mathcal{M}_G[i]$ is element-wise contained in the observed access vector $\mathbf{v}$:
\begin{equation}
\hat{\mathcal{C}} = \bigl\{\, i \in \{1,\dots,N\} \,:\, \mathbf{v}[j] \geq \mathcal{M}_G[i,j],\ \forall j \in \{1,\dots,4096\} \,\bigr\}.
\label{eq:filtering}
\end{equation}
Every true neighbor's footprint must appear in $\mathbf{v}$, so this step preserves all true neighbors, i.e., $\mathcal{C} \subseteq \hat{\mathcal{C}}$. However, $\hat{\mathcal{C}}$ may also contain unrelated nodes whose individual footprints happen to fall within the combined footprint of all true neighbors, inflating $\hat{\mathcal{C}}$ relative to $\mathcal{C}$. The subsequent steps use the estimated neighbor count $\hat{m}$ to remove these spurious candidates whenever doing so is reliable.

\subsection*{Step 2: Estimating $\hat{m}$ and Deriving the Trustworthiness Threshold $T$}

The neighbor count is estimated by dividing the total observed cache-line occupancy by the average per-node footprint:
\begin{equation}
\hat{m} = \frac{1}{\bar{f}} \sum_{j=1}^{4096} \mathbf{v}[j],
\label{eq:mhat}
\end{equation}
where $\bar{f}$ is the average number of cache lines occupied by one node's feature vector in the dataset (column \#Lines in Table~\ref{tab1}). The estimator $\hat{m}$ deviates from the true $m$ due to two error sources.

\noindent\emph{Discretization error.} All node feature vectors share the same dimension, but their starting offsets within cache lines vary, so each node occupies either $\lfloor \bar{f} \rfloor$ or $\lceil \bar{f} \rceil$ cache lines. Let $p = \bar{f} - \lfloor \bar{f} \rfloor$ denote the probability that a node occupies one extra line. Modeling the per-node extra-line indicator as i.i.d.\ Bernoulli, the total line count across $m$ neighbors has variance $m\,p(1-p)$, giving a standard deviation of
\begin{equation}
\sigma_{\mathrm{disc}}(m) = \sqrt{m \, p(1-p)}.
\label{eq:sigma-disc}
\end{equation}

\noindent\emph{Side-channel noise.} The second source of error stems from side-channel measurement inaccuracies. We empirically model this as a relative error coefficient $\eta$ applied to the total cache-line occupancy, so that the induced absolute error in $\sum_j \mathbf{v}[j]$ scales with $m\bar{f}$:
\begin{equation}
\sigma_{\mathrm{noise}}(m) = \eta \cdot m \cdot \bar{f}.
\label{eq:sigma-noise}
\end{equation}
From our experimental measurements, we estimate $\eta = 0.6\%$ for \textsc{CPrime}+\textsc{CProbe} and $\eta = 0.4\%$ for \textsc{GPrime}+\textsc{CProbe}, reflecting the cleaner SLC state produced by GPU-side priming.

Dividing both error terms by $\bar{f}$ converts them into deviations on $\hat{m}$:
\begin{equation}
\Delta_{\mathrm{disc}}(m) = \frac{\sqrt{m\,p(1-p)}}{\bar{f}}, \qquad \Delta_{\mathrm{noise}}(m) = \eta \cdot m.
\label{eq:delta}
\end{equation}

For $\hat{m}$ to round to the correct integer with at least $95\%$ confidence, the combined deviation must stay below $0.5$, where the (random) discretization term is scaled by the $95\%$ normal quantile $1.96$:
\begin{equation}
1.96 \cdot \Delta_{\mathrm{disc}}(m) \;+\; \Delta_{\mathrm{noise}}(m) \;<\; 0.5.
\label{eq:reliability}
\end{equation}
We define the \emph{trustworthiness threshold} $T$ as the largest $m$ at which~\eqref{eq:reliability} still holds, i.e., $T$ is the positive solution to
\begin{equation}
\boxed{\;1.96 \cdot \frac{\sqrt{T\,p(1-p)}}{\bar{f}} \;+\; \eta \cdot T \;=\; 0.5\;}
\label{eq:threshold}
\end{equation}
which is a quadratic equation in $\sqrt{T}$ and admits a closed-form solution. The resulting $T$ is dataset- and configuration-specific: a larger $\bar{f}$ or a smaller $p$ (more uniform per-node line counts) increases $T$, whereas a larger $\eta$ shrinks $T$. Because \textsc{GPrime} drives the SLC into a cleaner state than \textsc{CPrime}, \textsc{GPrime} has a smaller $\eta$ and therefore a larger $T$ on the same dataset. The two columns labeled ``Threshold $T$'' in Table~\ref{tab1} report the calibrated thresholds for each dataset under \textsc{CPrime} and \textsc{GPrime}.

When $\hat{m} < T$, the estimate is trustworthy and the pruning step (Step~3) may be safely applied. When $\hat{m} \geq T$, the estimate is unreliable and the procedure conservatively returns $\hat{\mathcal{C}}$ without pruning.

\subsection*{Step 3: Pruning}
Algorithm~\ref{alg:single-node} consumes $\hat{\mathcal{C}}$, $\hat{m}$, and $T$ and produces the final returned set $\tilde{\mathcal{C}}$. The procedure first checks whether $\hat{m}$ is trustworthy and whether pruning is necessary or computationally feasible. If pruning proceeds, it enumerates all size-$\hat{m}$ subsets of $\hat{\mathcal{C}}$ whose combined footprint is fully contained in $\mathbf{v}$, and consolidates the matching subsets by union so that any node appearing in at least one valid explanation is retained. In any non-actionable case (untrustworthy estimate, already-minimal candidate set, intractable enumeration, or no matching subset), the algorithm conservatively returns $\hat{\mathcal{C}}$ unchanged, preserving recall at the cost of some precision.
\begin{algorithm}[h]
\caption{Single-Node Edge Recovery: Pruning}
\label{alg:single-node}
\begin{algorithmic}[1]
\Require Candidate set $\hat{\mathcal{C}}$; estimate $\hat{m}$; access vector $\mathbf{v}$; threshold $T$; profiles $\mathcal{M}_G$; enumeration cap $K_{\max}\!=\!10^5$
\Ensure Returned neighbor set $\tilde{\mathcal{C}}$
\If{$\hat{m} \geq T$} \Comment{Estimate untrustworthy}
    \State \Return $\tilde{\mathcal{C}} \gets \hat{\mathcal{C}}$
\EndIf
\If{$\hat{m} \geq |\hat{\mathcal{C}}|$} \Comment{Candidate set already minimal}
    \State \Return $\tilde{\mathcal{C}} \gets \hat{\mathcal{C}}$
\EndIf
\State $K \gets \binom{|\hat{\mathcal{C}}|}{\hat{m}}$
\If{$K > K_{\max}$} \Comment{Enumeration intractable}
    \State \Return $\tilde{\mathcal{C}} \gets \hat{\mathcal{C}}$
\EndIf
\State $\mathcal{S}_{\mathrm{match}} \gets \emptyset$
\ForAll{subsets $S \subseteq \hat{\mathcal{C}}$ with $|S| = \hat{m}$}
    \If{$\mathbf{v}[j] \geq \sum_{i \in S} \mathcal{M}_G[i,j]\ \ \forall j$} \Comment{Combined footprint contained in $\mathbf{v}$}
        \State $\mathcal{S}_{\mathrm{match}} \gets \mathcal{S}_{\mathrm{match}} \cup \{S\}$
    \EndIf
\EndFor
\If{$|\mathcal{S}_{\mathrm{match}}| = 0$} \Comment{No matching subset}
    \State \Return $\tilde{\mathcal{C}} \gets \hat{\mathcal{C}}$
\ElsIf{$|\mathcal{S}_{\mathrm{match}}| = 1$}
    \State \Return $\tilde{\mathcal{C}} \gets$ the unique $S \in \mathcal{S}_{\mathrm{match}}$
\Else
    \State \Return $\tilde{\mathcal{C}} \gets \bigcup_{S \in \mathcal{S}_{\mathrm{match}}} S$ \Comment{Union of matching subsets}
\EndIf
\end{algorithmic}
\end{algorithm}
Because $\mathcal{C} \subseteq \hat{\mathcal{C}}$ holds by construction and every fall-back branch returns $\hat{\mathcal{C}}$, the procedure rarely drops true neighbors and therefore achieves near-100\% recall. Precision degrades on larger or denser graphs, where the combined footprint of many neighbors saturates the SLC, inflates $\hat{\mathcal{C}}$, and either pushes $\hat{m}$ above $T$ or makes the size-$\hat{m}$ enumeration intractable—both of which trigger the conservative fall-back to $\hat{\mathcal{C}}$.